\documentclass[12pt, a4paper]{book}
\usepackage[left=35mm,right=25mm,top=25mm,bottom=25mm]{geometry}

\usepackage{amsmath,amssymb,bm}
\usepackage{enumerate}
\usepackage{booktabs}

\usepackage{latexsym, algorithm, algorithmic}
\usepackage{hyperref}
\usepackage{graphicx}
\usepackage{ascmac}
\usepackage{wrapfig}

\renewcommand{\vec}{\bm}
\newcommand{\nvspace}[1]{

\vspace{#1\Cvs}

}

\newcommand{\FourDVec}[4]{\begin{pmatrix}#1 \\ #2\ \\ #3 \\ #4end{pmatrix}}

\newcommand{\whiteBox}{\hfill \ensuremath{\Box}}

\newcommand{\bra}{\langle}
\newcommand{\ket}{\rangle}

\begin{document}
\title{
\vspace{1.95cm}
\textbf{Scaling Concepts in Graph Theory:\\\vspace{12pt}Self-Avoiding Walk on \\Fractal Complex Networks} \vspace{10mm}}
\author{\textsc{Yoshihito Hotta}\\[13mm]\textsc{A Thesis}\\[2.5mm] Presented to the Faculty of the \\[1.5mm] \textsc{University of Tokyo}\\[2.5mm] In Candidacy for the Degree of  \\[2.5mm] \textsc{Master of Engineering}\\[25mm] \date{February 2014}}

\maketitle
\chapter*{Abstract}
%%%%%%%%%%%%%%%%%%%%%%%%%%%%%%%%%%%%%%
%       Nature形式のabstract
%%%%%%%%%%%%%%%%%%%%%%%%%%%%%%%%%%%%%%
% 1.ステージを設定
% 2. 同業者に向けた詳しい背景説明 (optional)
% 3. 問題点
% 4. 主結果を一文で述べる
% 5. ２，３文で分かったことを述べる。
% 6. より一般の文脈で分かったことや将来の展望を述べる
% 7. 全分野の科学者に向けて述べる（optional）
%%%%%%%%%%%%%%%%%%%%%%%%%%%%%%%%%%%%%%%
%%%%%%%%%%%%%%%%%%%%%%%%%%%%%%%%%%%%%%%
%\newpage
% 1.ステージを設定
It was discovered a few years ago that many networks in the real world exhibit self-similarity. 
% 2. 同業者に向けた詳しい背景説明 (optional)
A lot of researches on the structures and processes on real and artificial fractal complex networks have been done, drawing an analogy to critical phenomena. 
% 3. 問題点
However, the non-Markovian dynamics on fractal networks has not been understood well yet. 
% 4. 主結果を一文で述べる
We here study the self-avoiding walk on complex fractal networks through the mapping of the self-avoiding walk to the $n$-vector model by a generating function formalism. 
% 5. ２，３文で分かったことを述べる。
First, we analytically calculate the critical exponent $\nu$ and the effective coordination number (the connective constant) by a renormalization-group analysis in various fractal dimensions.
We find that the exponent $\nu$ is equal to the exponent of displacement, which describes the speed of diffusion in terms of the shortest distance.
Second, by obtaining an exact solution, we present an example which supports the well-known conjecture that the universality class of
the self-avoiding walk is not determined only by a fractal dimension. 
% 6. より一般の文脈で分かったことや将来の展望を述べる
Our finding suggests that the scaling theory of polymers can be applied to graphs which lack the Euclidian distance as well. 
Furthermore, the self-avoiding walk has been exactly solved only on a few lattices embedded in the Euclidian space, 
but we show that consideration on general graphs can simplify analytic calculations and leads to a better understanding of critical phenomena.
% 7. 全分野の科学者に向けて述べる（optional）
The scaling theory of the self-avoiding path will shed light on the relationship between path numeration problems in graph theory and statistical nature of paths. 

%\vspace{12pt}
%\begin{center}
%\begin{warning}
%Detailed explanation of my study
%\end{warning}
%\end{center}
%I calculated the critical exponent $\nu$ on a fractal complex network, so called the $(u,v)$-flower (H. D. Rozenfeld, S. Havlin, and D. Ben-Avraham, New J Phys 9, 175 (2007).
%). Here $u$ and $v$ are parameters of the graph. 
%For given two positive integers $u,v\ge 2$,
%\begin{align}
%\mu &= \frac{1}{x_{c}}\\
%\nu &= \frac{\ln(u)}{\ln\left( u x_{c}^{u-1} + v x_{c}^{v-1}\right)}\\
%\intertext{where $x_{c}$ is a positive fixed point of a scaling variable:}
%x'&=x^u + x^v
%\end{align}
%The fractal dimension of the $(u,v)$-flower is $d_{f} = \ln(u+v)/\ln u$.
%It can be shown that the range of $\nu$ is $0 < \nu \le 1$, the right equality holds true iif $u=v$, and $\nu$ can become arbitrarily close to $0$.
%\begin{figure}[b]
%\includegraphics[width=10cm,clip]{RG_flow_2}
%\label{fig:RG_flow_2}
%\centering
%\caption{An example of renormalization of a SAW path on $(2,2)$-flower. The decimation is carried away by erasing a smaller structure. }
%\end{figure} 

%\include{acknowledgements} 
\chapter*{Acknowledgement}
Foremost, I would like to show my greatest appreciation to Prof. Naomichi Hatano.
He gave me an opportunity to study statistical physics in my Master's course, and
he kindly taught me the theory of critical phenomena from A to Z.
In particular, the study of Monte-Carlo simulation could not be completed without his support.
I learned from him not only techniques, but also a phenomenological way of thinking and an attitude to pursue universality as a theoretical physicist.
He also read this long thesis and corrected many mistakes. 
%I could not have imagined having a better mentor for my graduate course.\\ 

Besides my advisor, I would like to acknowledge Tatsuro Kawamoto and Tomotaka Kuwahara for 
useful discussion. 
Their comments influenced this research a lot. 

I am also grateful to other members of the Hatano laboratory, Emiko Arahata, Masaaki Nakamura, Savannah Garmon, Youhei Morikuni, Hiroyasu Tajima, Masayuki Tashima, Rikugen Takagi, and Kaoru Yamamoto. 
I enjoyed chatting over coffee with them.

I thank Terufumi Morishita for advice on statistics and 
Liew Seng Pei for correcting mistakes in English of this manuscript.

Last but not least, I am indebted to Prof. Hidetoshi Katori.

\tableofcontents
\chapter{Introduction}
In this chapter, we briefly review previous studies and introduce the minimum amount of concepts 
which are required to understand this thesis. 
The thesis is made as self-contained as possible, 
but when there are elementary textbooks, we just cited them and 
avoided making the thesis lengthy.

First, we introduce the concept of the fractal. 
Because distinguishing various fractal dimensions is important to understand the fractality  of complex networks, we will define several fractal dimensions. 
Next, we review the basics of complex networks and present a model of fractal networks called the $(u,v)$-flower. 
We will consider the self-avoiding walk on the $(u,v)$-flower in the following chapters. 
Finally, we describe well-known conjectures on the self-avoiding walk and review the 
mapping of the self-avoiding walk on a graph to a zero-component ferromagnet.

\section{Fractal}
Structures that appear in nature are really rich in variety~\cite{mandelbrot83, falconer07, yakubo13}. 
Crystals possess discrete translational and rotational symmetries, and are classified by the point groups. 
On the other hand, molecules in gas and liquid are randomly distributed. Not all structures that appear in nature, however, are categorized to these two extreme classes. 
Many materials indeed fall in between these two classes;
 they partly possess a periodic structure and are partly random. 
Polymers, liquid crystals, and glasses are examples. 
If we consider objects in a wide sense, say, branching of trees, shapes of coastlines and rivers, and wrinkles of brains, 
most of them probably fall into the middle classes. 
When we discuss complexity, we do not say that objects with complete periodicity or complete randomness are complex; we regard objects which partly have both order and randomness as complex. 

\subsection{Fractal dimension}
Among the interesting properties of complex systems, a notable one is the self-similarity.
The self-similarity is a symmetry in which a part of a system is similar to the whole part. 
Of course we cannot expect that real objects in nature are self-similar in a mathematically rigorous sense, but many are so in a statistical sense. 
For instance, if we enlarge a picture of a ria coast, it will be as complex as the original picture is. 
If we are not told which is an enlarged one, we will not be able to answer which one is which. 
This means that a ria coast lacks a typical length. 
If there were a typical length, the ria coast would look completely different after magnification. On the other hand, if we magnify a picture of a coastline and the picture looks different when the picture is bigger than some size, it tells us that that size is the typical length of the coastline. 
Therefore, the self-similarity and lack of the typical length scale are equivalent.

When the `size' $M$ of an object is related with the `length' $L$ as
\begin{align}
M\propto L^{d_{\mathrm{f}}} ~~,             \label{eq:defFractalDim}
\end{align}
we say that the fractal dimension of that object is $d_{\mathrm{f}}$. There are many mathematically rigorous definitions, but we just write two definitions relevant to this thesis: the similarity dimension, the box-counting dimension, and the cluster dimension~\cite{falconer07}.

%the above definition is satisfactory because the the fractal dimension of $(u,v)$-flower is common for all the standard definitions of fractal definitions. 

\subsection{Similarity dimension}
Let us consider how to define a dimension of an object consisting of many small components. 
For instance, a cubic lattice is a collection of small cubes of edge length $l$. 
We use the smallest component as a unit to measure the `volume' of the whole object;
 we consider that the `volume' of the whole object is proportional to the number of the smallest components contained in the object.
 
%Next, let us use a bigger length $bl$ as a basic unit. Here, $b$ is an positive integer. 
%The number of these bigger components required to fill the whole object is smaller than that of smaller components needed to. Write $N(b)$ be the number of the smallest units, then
Let $N(b)$ be the number of the smallest components needed to fill a cube of edge length $L=bl$. 
We immediately see that $N(b)$ and $b$ are related as
\begin{align}
N(b) = b^{ 3 } 
\end{align}
in three dimensions. 
This is consistent with Eq. (\ref{eq:defFractalDim}).

Generalizing this argument, we want to define a dimension which is applicable to objects without the smallest unit, such as the Sierpinski gasket and the Cantor set. If an object of length scale $L$ consists of $b^{d_{\text{sim}}}$ pieces of objects of length scale $L/b$ , then we call $d_{\text{sim}}$ the similarity dimension. For instance, the Cantor set is created by deleting the middle open one third of a line segment repeatedly. 
Thus, the original set is restored by collecting two sets scaled down by $1/3$. As $2=3^{\log_{3}2}$, the similarity dimension of the Cantor set is $\log_{3}2$. 

The definition of the similarity dimension is applicable only to mathematical models, since fractals in nature possess the self-similarity only in a statistical sense.

\subsection{Box-counting dimension}
The similarity dimension is applicable only in limited cases as we explained. 
We would like to introduce another dimension which can be used more generally. 
We define such a dimension by borrowing the concept of the outer measure. 

Let the minimum number of cubes of edge length $l$ needed to cover an object be $N(l)$. 
If $N(l)$ and $l$ are related as
\begin{align}
N(l)\propto l^{-d_{\text{BC}}}  ,
\end{align}
we can measure the `volume' of the object because we know the volume of the cubes without ambiguity. We refer to $d_{\text{BC}}$ as the box-counting dimension. 
Precisely speaking, the box-covering dimension is defined as 
\begin{align}
d_{\text{BC}} = \lim_{l \searrow 0 }\frac{\log N(l)}{\log (1/l)}  . \label{eq:defBC}
\end{align}

Unlike the similarity dimension, the definition \eqref{eq:defBC} is directly applicable to fractals in nature as well as artificial fractals such as the Sierpinski gasket. 
It has indeed been known since long years ago that the length of a coastline depends on the precision of measurement. This reflects the fact that the fractal dimensions of coast lines are greater than unity.

\subsection{Cluster dimension}
As explained above, we can use the similarity dimension only for artificial fractals with a rigorous self-similarity. It would be convenient if the similarity dimension can be used for objects with a self-similarity in a statistical sense as the box-covering dimension.

Let us stipulate that a fractal has a minimum length scale. 
Let $\tilde{N}(L)$ be the average number of the minimum units inside a cube of edge length $L$.
As the similarity dimension is based on the number of the smaller units, 
we define a cluster dimension in terms of the average number of the minimum units:
\begin{align}
\tilde{N}(L)\propto L^{d_{\mathrm{c}}} . \label{eq:Ntilde}
\end{align}
We call $d_c$ the cluster dimension.
We can rephrase Eq. \eqref{eq:Ntilde} as
\begin{align}
\tilde{N}(L)=b^{d_{\mathrm{c}}} \tilde{N}(L/b) .
\end{align}

While there are many definitions of fractal dimensions, it is empirically known that fractal dimensions of fractals in nature seldom depend on the choice of the type of the fractal dimensionality. 
Hence, the definitions are usually not distinguished and just called `the fractal dimension $d_f$'. There are, however, cases where fractal dimensions strongly depend on the choice in complex networks as we will explain later.

\section{Complex network}
\subsection{Graph}
\begin{figure}
\includegraphics[width=15cm,clip]{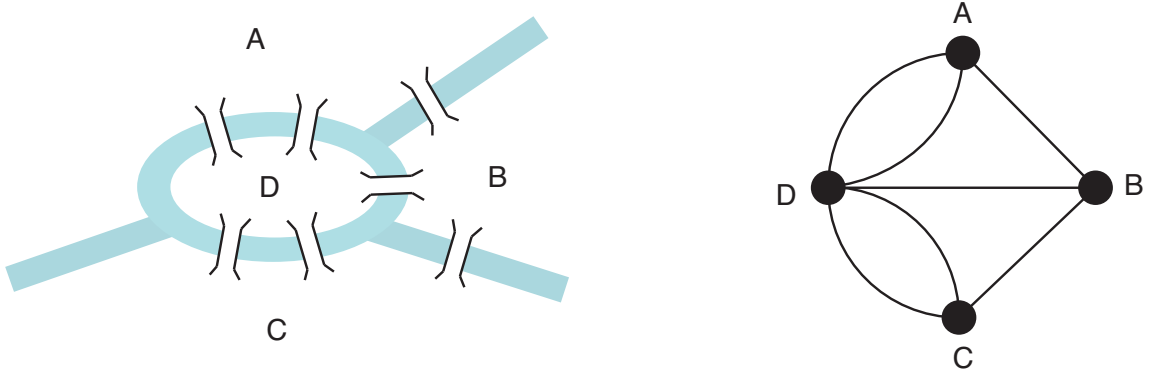}
\centering
\caption{The K\"onigsberg bridge problem. The city of K\"onigsberg has seven bridges across rivers (left). The problem is to find a route to pass every bridge in the city once and only once. 
The left figure can be abstracted into the right figure by replacing each land and bridge with a node and an edge, respectively.}
\label{fig:koenigsbergBridge}
\end{figure}

Graph theory has a long history. It began in the eighteenth century when a great mathematician Leonhard Euler visited K\"onigsberg. He asked himself whether there is a route
to visit every bridge in the city exactly once and to go back to the starting point (Figure \ref{fig:koenigsbergBridge}). The map of the city is originally a two-dimensional one, but in order to solve this problem we do not need the Euclidian distance; we can abstract the map. The abstracted map is represented by black circles and curves connecting the black circles. The black circles and curves are called nodes and edges, respectively. In this case, the nodes represent lands and the edges do bridges. 

The nodes and edges are not necessarily associated with physical objects; this kind of abstraction of problems is often useful. For instance, graphs often appear in problems of computer algorithms, which have clearly nothing to do with physical objects. 
For glossary of graph theory, refer to textbooks or web dictionaries~\cite{wilson10}.

%\subsection{Glossary of graph theory}
%Terms of graph theory which appear in this thesis are defined here. 
%\begin{enumerate}
%\item Graph\\
%A graph $G$ consists of two sets $V$ and $E$. $V$ is a discrete set and called the vertex set of $G$. $E$ is a subset of a direct product $V\times V$, and $E$ is called the edge set of $G$. 
%
%\item Node (vertex)\\
%A node is an element of vertex set $V$. A node is also called vertex.
%
%\item Edge\\
%An edge is an element of edge set $E$.
%
%\item Loop\\
%A loop is a form of edge $(v,v)\in E$.
%
%\item Walk\\
%A walk if a finite series of edges of the form $v_{0}\to v_{1}\to v_{2}\to\cdots \to v_{m}$.
%
%\item Path (simple path)\\
%A path is a walk of which nodes are all different except for the initial and the final nodes, i.e., 
%\begin{align}
%v_{0}\to v_{1}\to v_{2}\to\cdots \to v_{m},~~~~~v_{i}\neq v_{j}~~~(0< i < j<m)
%\end{align}
%A path is sometimes called simple path.
%
%\item Cycle\\
%A cycle is a path whose initial node and final node are same. A cycle is called 'node' in physics. We use 'loop' as a synonym of 'cycle' when there is no possibility of confusion.
%
%\item Connected graph\\
%Graph $G$ is called a connected graph when there exists a path between arbitrary two nodes of $G$. 
%
%\item Tree\\
%A graph which does node contain a cycle is defined as a forest. A connected forest is called tree. 
%
%\end{enumerate}

\subsection{Complex network}
Though there is no rigorous definition of complex networks, graphs which appear in real systems are usually called complex networks. 
The adjective complex is used because the real systems usually have a complex structure. 
Real networks possess both randomness and order to some extent. 
Their properties are different from Erd\"os-R\'enyi graphs, which are completely random, 
and at the same time different from periodic lattices~\cite{dorogovtsev03, watts99, barrat08,albert02, dorogovtsev02_1, newman03, dorogovtsev08, newman06, yakubo13, watts98, barabasi99}. 
Conditions of theorems of graph theory do not often hold in a rigorous sense, and hence we have to resort to some approximations.
 
Statistical physics has historically treated interactions of components that lie on a lattice with a translational symmetry and studied cooperative phenomena. Attempts to understand real materials have prompted physicists to develop numerous calculation techniques. Physicists have noticed that methodology of statistical mechanics is useful to understand networks, 
which have nothing to do with materials and had traditionally been thought to be outside the realm of physics. 

\subsection{Degree}
The number of edges connected to a node $i$ is called the degree of the node $i$ and denoted as $k_{i}$. Let $N$ be the total number of nodes and $M$ be the total number of edges.
We have
\begin{align}
\sum_{i=1}^{N} k_{i} = 2M  .
\end{align}
The average degree is 
\begin{align}
\bra k \ket = \frac{1}{N}\sum_{i=1}^{N} k_{i} = \frac{2M}{N}  .
\end{align}
We denote by $P(k)$ the probability that the degree of a randomly extracted node is $k$, 
which is called the degree distribution function. 
Using the degree distribution $P(k)$, we can rewrite the average degree as
\begin{align}
\bra k \ket = \sum_{k=0}^{\infty} kP(k) .
\end{align}
In many real networks, degree distributions are power functions as in $P(k)\propto k^{-a}$ with $a>0$, which is often called the scale-free property.

\subsection{Fractal complex networks}
Only fractals embedded in the Euclidian spaces have been studied until recently. 
In order to consider fractals in a space without the Euclidian distance, we have to 
introduce another distance 
because the fractal is fundamentally associated with the question as to ``how a volume grows as the system size increases".
There are many choices of a distance in graphs, but there is no standard choice as in the Euclidian space. 
Throughout this thesis, we use \textit{the shortest distance as the distance on graphs}.

The diameter of a graph is the largest shortest distance between all the pairs of two nodes. 
The mean shortest distance is the average over all pairs of nodes.
Let $N$ be the number of the nodes and $L$ be a diameter. The fractal dimension of the graph $d_{\mathrm{f}}$ may be intuitively given by
\begin{align}
N\propto L^{d_f} .
\end{align}
On the other hand, many real complex networks have a small-world property; the number of nodes and the mean shortest distances are related as 
\begin{align}
\bra l \ket \propto \log N .
\end{align}
Therefore, it seems that most of real complex networks are not fractals at a glance. \\

\begin{figure}
\includegraphics[width=15cm,clip]{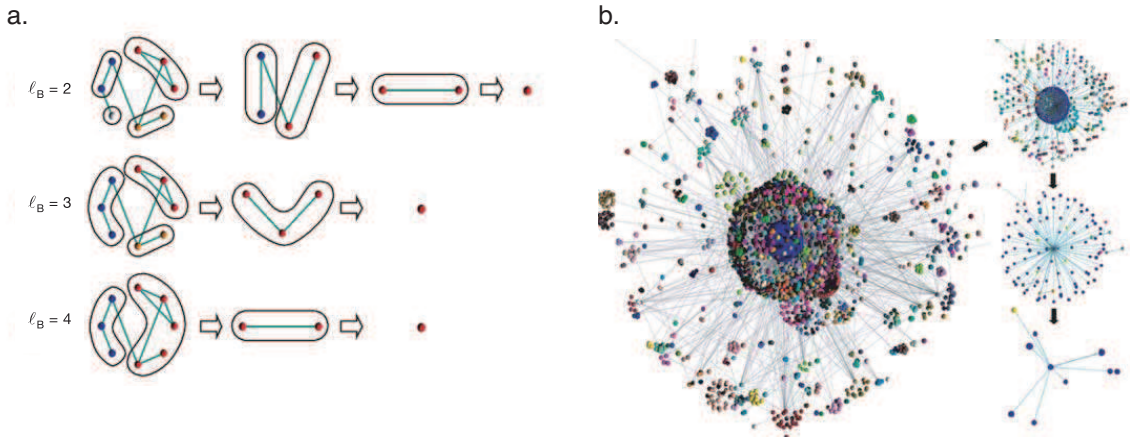}
\centering
\caption{The renormalization procedure applied to a real complex network.
{\bf a.} The box-covering method for a graph. We tile the graph with subgraphs whose diameter is
less than $l_{B}$. 
Then we replace each subgraph with a single node; two renormalized nodes are connected if there are at least one edge between the subgraphs. Thus we obtain the network shown in the second column. The decimation is repeated until the graph is reduced to a single node. 
{\bf b.} The renormalization is applied to the WWW network. The renormalized network is as complex as the original one. This indicates that the WWW network is a fractal. 
Taken from Song \textit{et al.}~\cite{song05}.
}
\label{fig:Song}
\end{figure}

Song \textit{et al.} found that a few graphs in real networks are indeed fractal (Figure $\ref{fig:Song}$)~\cite{song05,gallos07a}. 
They noticed that complex networks that had been studied many times were fractal, {\it i.e.}, 
\begin{enumerate}
\item a part of the WWW composed of 325,729 web pages, which are connected if there is a URL link from one page to another; 
\item a social network where the nodes are 392,340 actors, who are linked if they were cast together in at least one film;
\item the biological networks of protein-protein interactions found in \textit{Escherichia coli} and \textit{Homo sapiens}, where proteins are linked if there is a physical binding between them. \\
\end{enumerate}

Song \textit{et al.} argued that because of the long-tail distribution of degrees of nodes the cluster dimension $d_{\mathrm{c}}$ and $d_{\text{BC}}$ are not identical in scale-free networks.

\subsection{The $(u,v)$-flower}
\begin{figure}
\includegraphics[width=13cm,clip]{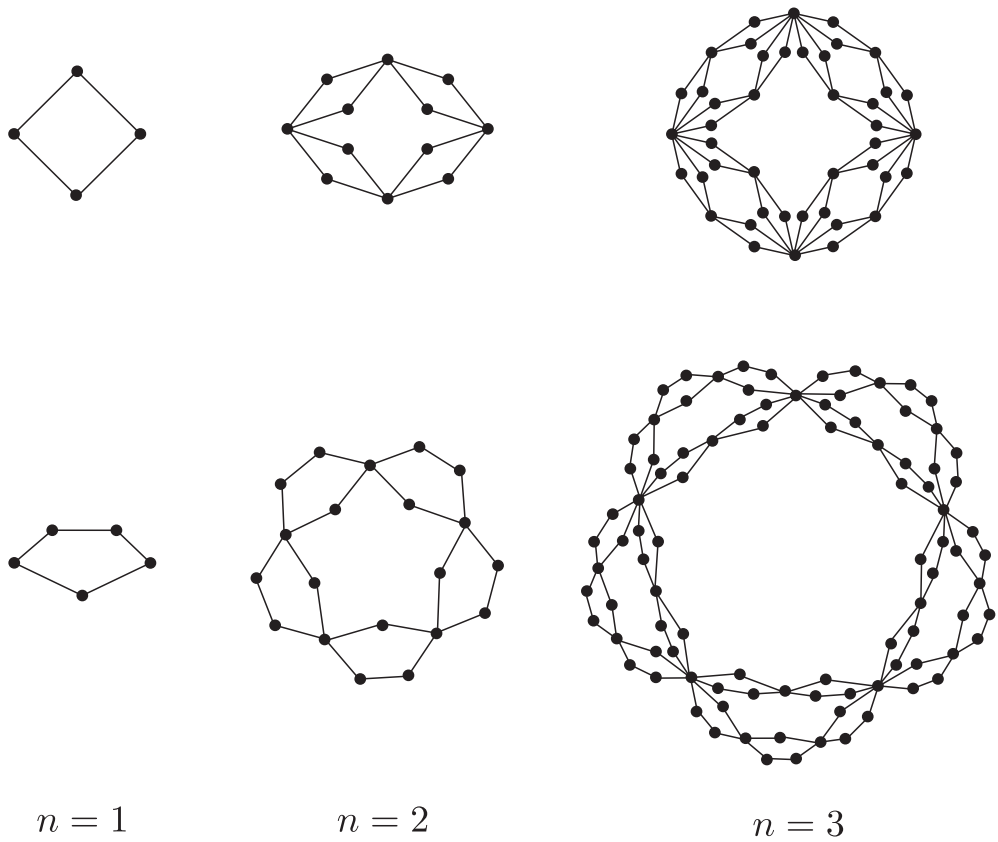}
\centering
\caption{The $(2,2)$-flower and the $(2,3)$-flower in the first, second and third generations. Each line is replaced by parallel lines of length $u$ and $v$ in construction of the next generation.
}
\label{fig:uv_flower}
\end{figure}

After the discovery in the real networks, several artificial fractal complex networks have been devised~\cite{song06}. 
One of such networks is the $(u,v)$-flower (Figure $\ref{fig:uv_flower}$)~\cite{rozenfeld07_1}. 
As deterministic fractals such as the Sierpinski gasket and the Cantor set helped us understand real fractals in the Euclidian spaces, 
deterministic fractal complex networks can deepen our understanding of fractal complex networks in the real world. 
As with many other artificial fractals, the $(u,v)$-flower is a graph with a hierarchical structure~\cite{dorogovtsev02_2, dorogovtsev02_1}. 

The $(u, v)$-flower is defined in the following way. First, we prepare a cycle of length $u+v$ as the first generation. Second, given a graph of generation $n$, we obtain the $(n+1)$th generation by replacing each link by two parallel paths of length $u$ and $v$. We can assume $1\le u\le v$ without losing generality. 

Let $M_n$ and $N_n$ be the numbers of edges and nodes, respectively. From the definition of the $(u,v)$-flower, it straightforwardly follows that
\begin{align}
M_n &= w^n ,\\
N_n &= w N_{n-1} - w = \cdots = \frac{w-2}{w-1}\times w^n + \frac{w}{w-1} ,
\intertext{where}
w&=u+v .
\end{align}
The mean degree of $(u,v)$-flower in the $n$th generation is
\begin{equation}
\bra k \ket = \frac{2M_n}{N_n} .
\end{equation}

Similar consideration tells us about the degree distribution. The $(u,v)$-flowers only have nodes of degree $k=2^{m}$, where $m=1,2,\cdots,n$. Let $N_{m}(m)$ be the number of nodes of degree $2^{m}$ in the $n$th generation.
We thereby have
\begin{align}
N_{n}(m) = 
	\begin{cases}
	N_{n-1}(m-1) &\text{for~} m>1,\\
	(w-2)w^{n-1} &\text{for~} m=1.
	\end{cases}
\end{align}
Solving this recurrence relation under the initial condition $N_1(1)=w$, we have
\begin{equation}
N_n(m) = 
\begin{cases}
	(w-2)w^{n-m} &\text{for~} m<n,\\
	w     &\text{for~}m=n ,
\end{cases}
\end{equation}
which is related to the degree distribution $P(k)$ in the form $|N_n(m)dm| = |P(k)dk|$. 
We therefore have the degree distribution of the $(u,v)$-flower with $u,v\ge 1$ as 
\begin{equation}
P(k)\propto k^{-\gamma}~~\text{with}~\gamma=1+\frac{\ln(u+v)}{\ln 2} .
\end{equation}

The dimensionality of the $(u,v)$-flowers is totally different for $u=1$ and $u>1$~\cite{yakubo13}. When $u=1$ the diameter $d_n$ of the $n$th generation is proportional to the generation $n$, while the diameter $d_n$ is a power of $u$ when $u>1$:
\begin{equation}
d_n \sim 
\begin{cases}
	(v-1)n &\text{~for~} u=1 ,\\
	u^n    &\text{~for~} u>1 .
\end{cases} \label{eq:dnun} 
\end{equation}
Since $N_n\sim w^n$, we can transform Eq. \eqref{eq:dnun} to
\begin{equation}
d_n\sim
\begin{cases}
	\ln N_n &\text{~for~} u=1, \\
	N_n^{\ln u/\ln(u+v)} &\text{~for~}u>1 .
\end{cases}
\end{equation}
This means that the $(u,v)$-flowers have a small-world property only when $u=1$, while the flowers have finite fractal dimensions for $u>1$.

When $u>1$, it is clear from the construction of flowers that the similarity dimension of  the $(u,v)$-flower is
\begin{equation}
d_{\text{sim}} = \frac{\ln(u+v)}{\ln u} \text{~for~} u>1   .
\end{equation}
Because the cluster dimension is an extension of the similarity dimension, the cluster dimension of the $(u,v)$-flower is the same as that of the similarity dimension for $u>1$:
\begin{equation}
d_{\mathrm{c}} = d_{\text{sim}} = \frac{\ln(u+v)}{\ln u} \text{~for~} u>1  .
\end{equation}

\section{Self-avoiding walk}
A self-avoiding path, which is called a simple path or just a path in graph theory, is a path on a lattice (graph) that is forbidden to visit the same point more than once~\cite{flory53}. 
This path is referred to as the self-avoiding path throughout this thesis in order to distinguish it from other stochastic processes. 

Though the definition is quite easy, 
many important questions are still open in the Euclidian spaces even today~\cite{madras96}.
For example,
\begin{enumerate}
\item How many possible self-avoiding paths of length $k$ are there?
\item How long is the typical distance from the starting point?
\end{enumerate}
The goal of this thesis is to find a graph on which these questions are answered.

\subsection{Self-avoiding walk in a Euclidian space}
In a Euclidian space, the number of paths of length $k$, which is written as $C_{k}$, on $\mathbb{R}^{n}$ is believed to behave as
\begin{equation}
C_{k}\sim \mu^{k}k^{{\gamma-1}} \label{eq:CkEuclidian}
\end{equation}
and the mean square distance of paths of length $k$, which is denoted as $\bra R_{k}^{2}\ket$, is hypothesized to be
\begin{equation}
\bra R_{k}^{2}\ket \sim k^{{2\nu}}.
\end{equation}
Here the sign $\sim$ denotes the asymptotic form of the function as $k\to\infty$. 
The constant $\mu$ is called the connective constant, which roughly means the effective coordination number. The exponent $\gamma$ is a critical exponent associated with the susceptibility and $\nu$ is one associated with the correlation length from the viewpoint of the correspondence between the self-avoiding walk and the $n$-vector model. Thus, $\mu$ is sensitive to the specific form of the lattice, while $\gamma$ and $\nu$ are universal quantities, that is,  they are insensitive to the specific form of the lattice and are believed to depend only on the Euclidian dimension. 
The critical exponents are conjectured to be
\begin{align}
\gamma = \begin{cases}
\displaystyle\frac{43}{32} &\text{~for~}d=2  ,\\
1.162\dots &\text{~for~}d=3 ,\\
1~~\text{with a logarithmic correction} &\text{~for~}d=4 ,\\
1 &\text{~for~}d=5 ,
\end{cases}
\end{align}
\begin{align}
\nu = \begin{cases}
\displaystyle\frac{3}{4} &\text{~for~}d=2 ,\\
0.59\dots &\text{~for~}d=3 ,\\
1/2~~\text{with a logarithmic correction} &\text{~for~}d=4 ,\\
1/2 &\text{~for~}d=5.
\end{cases}
\end{align}
The upper critical dimension of the self-avoiding walk is $d=4$, 
above which the critical exponents are given by a mean-field model. 
The mean-field model of the self-avoiding walk is the random walk, 
whose critical exponent $\nu$ is $1/2$ as is well known.\\
 
Going beyond the Euclidian dimension, the self-avoiding walk in fractal dimensions has also been actively studied since the 1980s. It has been conjectured that the universality class of the self-avoiding walk of fractals are {\it not} determined just by a fractal dimension (precisely speaking the similarity dimension). 
Physicists have tried to express the exponent $\nu$ by the similarity dimension as an extension of Flory's approximation in the Euclidian space~\cite{rammal84, havlin87, aharony89}. 
\begin{align}
\nu=\frac{3}{2+d}~~~\longrightarrow~~~\nu=\frac{3}{2+d_{\text{sim}}}  .
\end{align}
They, however, found that replacement of the Euclidian dimension of Flory's approximation with the similarity dimension sometimes gives a deteriorated accuracy.
It was concluded that there is no simple formula for a fractal as in the Euclidian space.

\subsection{$n$-vector model}
This subsection describes the correspondence between the self-avoiding walk and 
a zero-component ferromagnet. 
The connection was first discovered by de Gennes~\cite{deGennes72, deGennes79}, and opened a way to study
a polymer in terms of the standard theory of critical phenomena. 
Shapiro~\cite{shapiro78} introduced a generating function, whose divergence near a pole governs 
the behavior of the zero-component ferromagnet at the critical point. 
We here follow the discussion by Madras and Slade~\cite{madras96}. 
Their mapping of the self-avoiding walk to the $n$-vector model is straightforward 
and can be directly applied to graphs as well as usual lattices. 

Assume that spins are on a graph $G=(V,E)$. 
The spins have $n$ components and the tip of each spin is on a sphere of radius $\sqrt{n}$:
\begin{align}
\vec{S}^{(x)} &= (S^{(x)}_{1}, S^{(x)}_{2}, \cdots, S^{(x)}_{n}) \in \mathcal{S}(n,\sqrt{n}) ,
\end{align}
where $\mathcal{S}(m,r)$ is the sphere of radius $r$ in $\mathbb{R}^{m}$:
\begin{align}
\mathcal{S}(m,r) &= \{(a_{1},a_{2},\cdots,a_{m})\in \mathbb{R}^{m} : a_{1}^{2} + a_{2}+ \cdots + a_{m}^{2}=r^{2}\}. 
\end{align}

We consider the Hamiltonian with a ferromagnetic interaction given by
\begin{align}
H = -\sum_{\bra x,y\ket} \vec{S}^{(x)}\cdot\vec{S}^{(y)} ,
\end{align}
where $x$ and $y$ are nodes, and $\bra x,y\ket$ is the edge connecting $x$ and $y$. The sum runs over all edges. 
The expectation value of any quantity $A$ is 
\begin{align}
\bra A \ket &= \frac{1}{Z} E(Ae^{-\beta H})
\intertext{with}
Z &= E(e^{-\beta H})  ,\label{eq:Znvec}
\end{align}
where $E(\cdot)$ is the expectation value with respect to the product of the uniform measure on $\mathcal{S}(n,\sqrt{n})$. 

The quantity of our interest is the correlation function in the limit $n\to 0$:
\begin{equation}
\lim_{n\to 0} \bra \vec{S}_{i}^{(x)}\cdot\vec{S}_{j}^{(y)} \ket .
\end{equation}
The limit $n\to 0$ is an extrapolation and not a mathematically justified procedure. 
We therefore have to explain its meaning. 
The limit should be defined so as to be consistent with the following lemma~\cite{madras96}:\\

\textit{Fix an integer $n\ge 1$. Let $\vec{S} = (S_{1},S_{2},\cdots,S_{n})$ denote a vector 
which is uniformly distributed on $\mathcal{S}(n,\sqrt{n})$. Given nonnegative integers $k_{1},\cdots,k_{n}$, }
\begin{align}
E(S_{1}^{k_{1}}S_{2}^{k_{2}}\cdots S_{n}^{k_{n}}) = 
\begin{cases}
\frac{2\Gamma\left( \frac{n+2}{2}\right)\prod_{l=1}^{n}\Gamma\left( \frac{k_{l}+1}{2}\right)}{\pi^{n/2}\Gamma\left( \frac{k_{1}+\cdots+k_{n}+n}{2}\right)} n^{(k_{1}+\cdots+k_{n}-2)/2} ~~~~~~& \text{\textit{when all $k_{l}$ are even}},\\
0 ~~~~~& \text{\textit{otherwise}}.
\end{cases}
\end{align}
\whiteBox 

We can prove it by mathematical induction.\\

We define the limit $n\to0$ in the following way. 
First, the following trivial equality holds:
\begin{equation}
E(1)=1  .
\end{equation}
Second, since $E(S_{1}^{2}+\cdots+S_{n}^{2})=n$, it follows from the symmetry that 
\begin{equation}
E(S_{i}^{2})=1  .
\end{equation}
Third, when $k_{1}+\cdots + k_{n}>2$, the exponent of $n^{(k_{1}+\cdots+k_{n}-2)/2}$ is positive. 
For these three reasons, we define the limit $n\to 0$ as follows.
\begin{align}
\lim_{n\to 0} E(S_{1}^{k_{1}}S_{2}^{k_{2}}\cdots S_{n}^{k_{n}}) = 
\begin{cases}
1 ~~~~~&\text{all $k_{l}=0$, or one $k_{l}=2$ and $k_{j}=0~~(j\neq l)$} ,\\
0~~~~~&\text{otherwise} .
\end{cases} \label{eq:limOfN}
\end{align}

In order to evaluate Eq. \eqref{eq:Znvec}, 
we expand the Boltzmann factor as the following power series:
\begin{align}
e^{-\beta H} = \prod_{\bra x,y\ket}\exp[ \beta \vec{S}^{(x)}\cdot\vec{S}^{(y)} ]
= \prod_{\bra x,y\ket} \sum_{m_{xy}=0}^{\infty} \frac{\beta^{m_{xy}}}{m_{xy}!} (\vec{S}^{(x)}\cdot\vec{S}^{(y)})^{m_{xy}}. \label{eq:ZnvecPower}
\end{align}
Let us label the edges as $e_{1},\cdots,e_{|E|}$. In this notation, Eq. \eqref{eq:ZnvecPower} can be rewritten as
\begin{align}
e^{-\beta H} = \sum_{m_{1},\cdots, m_{|E|}=0}^{\infty} \frac{\beta^{\sum_{\alpha\in E}m_{\alpha}}}{\prod_{\alpha\in E}m_{\alpha}!} \prod_{\alpha\in E} ( \vec{S}^{(e_{\alpha}^{-})}\cdot \vec{S}^{(e_{\alpha}^{+})} )^{m_{\alpha}}   .
\end{align}

Consider now the partition function
\begin{align}
Z =  \sum_{m_{1},\cdots, m_{|E|}=0}^{\infty} \frac{\beta^{\sum_{\alpha\in E}m_{\alpha}}}{\prod_{\alpha\in E}m_{\alpha}!}E\left( \prod_{\alpha\in E} ( \vec{S}^{(e_{\alpha}^{-})}\cdot \vec{S}^{(e_{\alpha}^{+})} )^{m_{\alpha}} \right)  .   \label{eq:partitionNModel}
\end{align}
\begin{figure}
\includegraphics[width=7cm,clip]{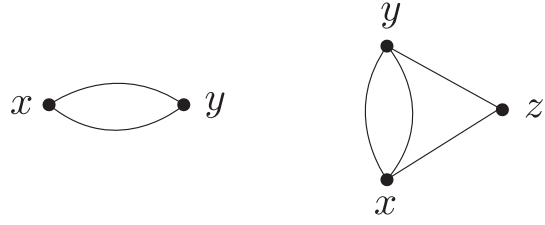}
\centering
\caption{Examples of the graphical representation of terms in Eq. (\ref{eq:partitionNModel}). 
The left diagram corresponds to $E( (\vec{S}^{(x)}\cdot\vec{S}^{(y)})^{2} )$ and is called a two-edge polygon. 
The right diagram represents $E((\vec{S}^{(x)}\cdot\vec{S}^{(y)})^{2}  (\vec{S}^{(y)}\cdot\vec{S}^{(z)}) (\vec{S}^{(x)}\cdot\vec{S}^{(z)}))$ .
}
\label{fig:selfAvoidingPolygon}
\end{figure} 
A graphical interpretation of the sum in Eq. (\ref{eq:partitionNModel}) can be obtained by associating to each term in the sum a graph whose each edge $e_{\alpha}$ is duplicated $m_{\alpha}$ times (if $m_{\alpha}=0$, then it means that the edge is removed) (Figure \ref{fig:selfAvoidingPolygon}). 
It follows from Eq. (\ref{eq:limOfN}) that any term whose corresponding graph has a node from which other than two or zero edges emanate will approach zero in the limit as $n\to 0$. 
Therefore, the only terms which may contribute are one with no edges and ones with self-avoiding polygons. 

A two-edge polygon with nearest-neighbor nodes $x,y$ (Figure \ref{fig:selfAvoidingPolygon}, left) contributes the amount 
\begin{equation}
\frac{\beta^{2}}{2}E( (\vec{S}^{(x)}\cdot\vec{S}^{(y)})^{2} ) = \frac{\beta^{2}}{2} N.
\end{equation}
Thus, a two-edge polygon is irrelevant in the limit $n\to 0$. A non-degenerate polygon, in other words a polygon consisting of at least three edges, also does not contribute according to a similar argument. 
The only term which is relevant in Eq. (\ref{eq:partitionNModel}) is a graph with no edges.
We therefore have 
\begin{align}
\lim_{n\to 0}Z = 1  .  \label{eq:partitionNModelToZero}
\end{align}

For the correlation function, the analysis is similar. 
We would like to compute the limit $n\to 0$ of the expectation value for $x\neq y$:
\begin{align}
\sum_{m_{1},\cdots, m_{|E|}=0}^{\infty} \frac{\beta^{\sum_{\alpha\in E}m_{\alpha}}}{\prod_{\alpha\in E}m_{\alpha}!}E\left( S_{i}^{(x)} S_{j}^{(y)}\prod_{\alpha\in E} ( \vec{S}^{(e_{\alpha}^{-})}\cdot \vec{S}^{(e_{\alpha}^{+})} )^{m_{\alpha}} \right)  .
\end{align}
Terms corresponding to graphs with self-avoiding polygons do not contribute because of the same reason. 
The only surviving terms are ones with self-avoiding paths from $x$ to $y$. 
Contribution due to the self-avoiding path $(x,v_{1},\cdots,v_{k-1},y)$ is 
\begin{align}
\beta^{k} E(S_{i}^{(x)} (\vec{S}^{(x)} \cdot \vec{S}^{(v_{1})}) (\vec{S}^{(v_{1})} \cdot \vec{S}^{(v_{2})}) \cdots (\vec{S}^{(v_{k-1})} \cdot \vec{S}^{(y)})  S_{j}^{(y)} )
= \beta^{k} \delta_{i,j}  .
\end{align}

All the contributing terms can be summed using the  
generating function of the $s-t$ paths connecting nodes $s$ and $t$:
\begin{align}
G_{z}(s,t) := \sum_{\omega:s\to t} z^{|\omega|}   .
\end{align}
Here $\omega$ is a simple path from $s$ to $t$, and $|\omega|$ denotes the length of the path $\omega$. 
The generating function $G_{z}(s,t)$ is often called the two-point function.

Using the generating function $G_{z}(x,y)$ and (\ref{eq:partitionNModelToZero}), 
we have
\begin{align}
\lim_{n\to 0} \bra \vec{S}_{i}^{(x)}\cdot\vec{S}_{j}^{(y)} \ket 
&= \sum_{m_{1},\cdots, m_{|E|}=0}^{\infty} \frac{\beta^{\sum_{\alpha\in E}m_{\alpha}}}{\prod_{\alpha\in E}m_{\alpha}!}E\left( S_{i}^{(x)} S_{j}^{(y)}\prod_{\alpha\in E} ( \vec{S}^{(e_{\alpha}^{-})}\cdot \vec{S}^{(e_{\alpha}^{+})} )^{m_{\alpha}} \right) \\
&= \delta_{i,j} \sum_{\omega:x\to y} \beta^{|\omega|}
= \delta_{i,j}G_{\beta}(x,y) .
\end{align}
Now, the correspondence between the self-avoiding walk and the zero-component ferromagnet is  established:

%\begin{screen}
\begin{align}
\lim_{n\to 0} \bra \vec{S}_{i}^{(x)}\cdot\vec{S}_{j}^{(y)} \ket 
= \delta_{i,j}G_{\beta}(x,y)  .
\end{align}
%\end{screen}
This relation holds on any graphs as well as on usual lattices. 

\section {Outline of this thesis}
In the following chapters, we consider the self-avoiding walk on the $(u,v)$-flower for $u,v\ge 2$. 

In Chapter 2, we first address analytic results based on the generating function formalism.  
We will derive the critical exponent $\nu$ and the connective constant $\mu$ under several assumptions. 
The most important result of this chapter is to present a counterexample which shows that there is no one-to-one correspondence between the universality class of the self-avoiding walk and the similarity dimension. 

In Chapter 3, we will confirm the assumptions used in Chapter 2 by numerical simulations. 
We introduce an enumeration algorithm and a Monte-Carlo algorithm which can be used for the self-avoiding walk on graphs lacking the translational and rotational symmetries. 
We will define the ensemble of paths of fixed length and consider the mean shortest end-to-end distance in that ensemble. 
We observe that the mean shortest end-to-end distance increases as in $\overline{d_{k}^{(s)}} \approx k^{\nu'}$. 
Furthermore, we will see that $\nu=\nu'$. 

In Appendices, we will explain additional results of numerical simulations and the detail of analysis in the Monte-Carlo simulation.

\chapter{Analytic Results}
In the previous chapter, we mentioned that the complex network $(u,v)$-flower is a fractal for $u,v\ge2$. We assume $2\le u \le v$ from now on. 
We delve into the self-avoiding walk on the $(u,v)$-flower in this chapter. 
We extend the theory of the self-avoiding walk in the Euclidian spaces and fractals~\cite{hattori04, shapiro78, dhar78,rammal84} to the $(u,v)$-flowers and formulate 
the exact renormalization of a two-point function between two hubs. 
We thereby derive analytic expressions of the connective constant $\mu$ and the critical exponent $\nu$, 
and thus determine the universality class of the self-avoiding walk in various fractal dimensions $1 < d_{f} < \infty$. 
Our result confirms the well known conjecture that there is no one-to-one correspondence between the fractal dimension and a critical exponent~\cite{gefen80, rammal84, havlin87, aharony89}. 
This means that unlike in the Euclidian space, the self-avoiding walk in fractals cannot be categorized to a few universality classes, but there exist the infinite number of classes.

\section{Renormalization of a propagator between hubs}
\begin{figure}
\includegraphics[width=13cm,clip]{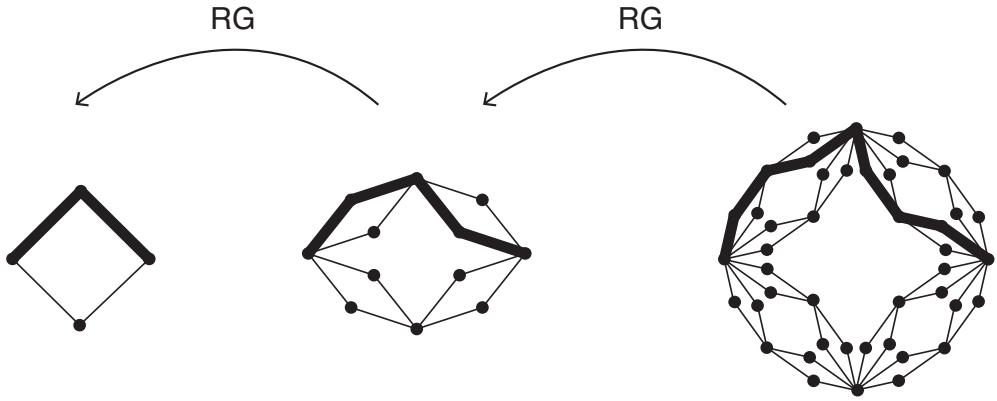}
\centering
\caption{An example of renormalization of a self-avoiding path on the $(2,2)$-flower. The decimation is carried out by erasing a smaller structure. }
\label{fig:RG_flow_2}
\end{figure}

We define a renormalization procedure for the $(u,v)$-flower as the inverse transformation of the constructing procedure of the flower (Figure $\ref{fig:RG_flow_2}$). 
When the $(n+1)$th generation is given, seeing the graph from a distance, we neglect the minute structure and obtain the $n$th generation. Every cycle of length $(u+v)$ is, therefore, replaced by a single edge. 
Renormalization of self-avoiding paths is also defined in a similar way. 

Let $R$ be a node which is away from a node $O$ in the first generation, and $R_{n}$ be \textit{the shortest distance} between $O$ and $R$ in the $n$th generation. 
The nodes $O$ and $R$ have the largest degree and are called hubs. 
Because each edge is replaced by two parallel lines of length $u$ and $v$ in the construction, 
$R_{n}$ increases as 
\begin{align}
R_{n} = R_{n-1}\times u = \cdots = u^{n-1}R_{1} = u^{n}
\end{align}

Defining $C_{k}^{(n)}(R)$ as the number of self-avoiding paths of length $k$ starting from the node $O$ and ending at the node $R$ in the $n$th generation, 
we can construct the two-point function as 
\begin{equation}
G_{n}(R_{n}, x) = \sum_{k=1} C_{k}^{(n)}(R) x^{k}   .\label{eq:genFunc}
\end{equation}
Let us assume that $C_{k}^{(n)}(R)$ behaves asymptotically as 
\begin{equation}
C_{k}^{(n)}(R)^{1/k}\sim \mu,
\end{equation}
because at each step a walker has $\mu$ options to go next on average. 
Then the convergence disk of (\ref{eq:genFunc}) is $|x|<1/\mu=:x_{c}$, where $x_{c}$ is a critical point.

The two-point function of the first generation is
\begin{align}
G_{1}(R_{1},x) = x^{u} + x^{v}
\end{align}
by definition. 
Since the $(n+1)$th generation can be regarded as a cycle of $(u+v)$ pieces of the $n$th generation graphs, 
\begin{align}
G_{n+1}(R_{n+1}, x)=G_{n}(R_{n}, x)^{u} + G_{n}(R_{n}, x)^{v}  .\label{eq:recursionG_n}
\end{align}
Therefore,
\begin{align}
G_{n+1}(R_{n+1}, x) = G_{1}(R_{1}, G_{n}(R_{n}, x)) .
\end{align}
Repeated use of this relation yields
\begin{align}
G_{n}(R_{n}, x) = \underbrace{G_{1}\circ G_{1}\circ\cdots \circ G_{1}}_{n} (R_{1},x)  .\label{eq:G_1ntimes}
\end{align}

\section{Renormalization-group analysis} \label{sec:RG}
The mapping of the self-avoiding walk to the $n$-vector model suggests that the two-point function becomes in the thermodynamic limit 
\begin{align}
G_{n}(R_{n},x) &\sim \exp(-R_{n}/ \xi(x)) \text{~~as~~}n\to\infty,
\intertext{where $\xi(x)$ is the correlation length, which should behave as}
\xi(x)&\sim (x_{c}-x)^{-\nu}~~~(x \nearrow x_{c})  .  \label{eq:nuDef}
\end{align}

\begin{figure}
\includegraphics[width=10cm,clip]{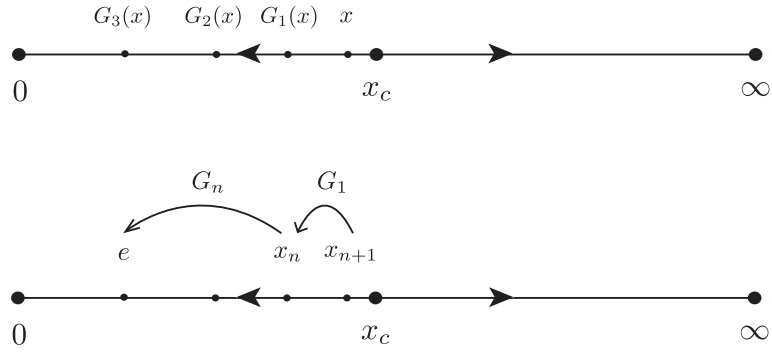}
\centering
\caption{The renormalization-group flow. The top figure illustrates how $G_{n}(x)$ changes as the generation $n$ gets larger with $x$ fixed. The bottom figure shows the flow of the scaling variable $x$. Here, $x_{n+1}$ is the scaling variable in the $(n+1)$th flower, and $x_{n}$ is the one in a coarse-grained flower.
 }
 \label{fig:RG_flow_1}
\end{figure} 

The critical exponent $\nu$ may be obtained by studying $\xi(x)$ near a fixed point. 
In the original problem, we wanted to study the asymptotic behavior as $x\to x_{c}$ for a fixed $n$.
In renormalization, we study how scaling variable $x$ changes as we perform the scaling transformation, rather than directly moving $x$ close to $x_{c}$ (Figure $\ref{fig:RG_flow_1}$). 
Let $e$ be a sufficiently small positive number. We define the variable $x_{n}$ such that
\begin{equation}
G_{n}(R_{n},x_{n}) := e  ~~\text{for all~}n.  \label{eq:defOfx_n}
\end{equation}
The $x_{n}$ is the scaling variable of our theory, and we observe how it transforms under 
the renormalization transformation. 
We will prove the unique existence of $x_{n}$ which satisfies Eq. \eqref{eq:defOfx_n} later. 

\begin{figure}
\includegraphics[width=13cm,clip]{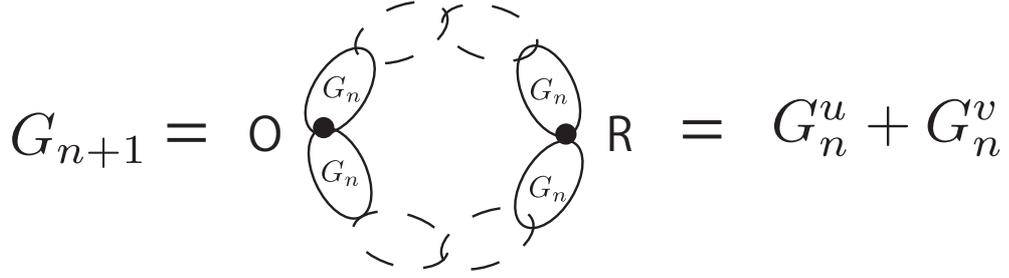}
\centering
\caption{Diagrammatic representations of the two-point function $G_{n}(R_{n},x)$. 
The $(n+1)$th generation can be regarded as a cycle of $(u+v)$ pieces of graphs in the $n$th generation.
}
\label{fig:diagram}
\end{figure}

The two-point function of the $(n+1)$th generation and that of the $n$th generation are related as
\begin{equation}
G_{n+1}(R_{n+1},x_{n+1}) = e = G_{n}(R_{n},x_{n})   .\label{eq:G_np1G_n}
\end{equation}
This specifies how the scaling variable $x$ is renormalized.
From (\ref{eq:G_1ntimes}) and (\ref{eq:G_np1G_n}), we obtain
\begin{align}
G_{n}(R_{n},x_{n}) = G_{n+1}(R_{n+1},x_{n+1}) = G_{n}(R_{n}, G_{1}(R_{1}, x_{n+1}) ) = G_{n}(R_{n}, x_{n+1}^{u} + x_{n+1}^{v})  .
\end{align}
The scaling variable therefore changes under the renormalization transformation as in
\begin{align}
x_{n} = x_{n+1}^{u} + x_{n+1}^{v}   .\label{eq:RG_flow}
\end{align}
We will later show that the scaling variable $x_{n}$ changes as in Figure \ref{fig:RG_flow_1}. 
The two-point function $G_{n}(R_{n}, x)$ and the scaling variable $x_{n}$ are transformed in the opposite ways (Figure $\ref{fig:diagram}$). 

Near a fixed point, 
\begin{align}
 \frac{R_{n+1}}{\xi(x_{n+1})} &= \frac{R_{n}}{\xi(x_{n})} ,
\end{align}
and hence
\begin{align}
(x_{c} -x_{n+1} )^{-\nu}& \sim \frac{R_{n+1}}{R_{n}} (x_{c} -x_{n} )^{-\nu} = u (x_{c} -x_{n} )^{-\nu} 
\end{align}
in the limit $n\to\infty$. 
The critical exponent $\nu$ is therefore expressed as 
\begin{align}
\nu = \frac{\ln(u)}{\ln\left( \frac{x_{c} - x_{n}}{x_{c}-x_{n+1}}\right)} = \frac{\ln(u)}{\ln\left( \frac{x_{n} - x_{c}}{x_{n+1}-x_{c}}\right)}   . \label{eq:nuTwoGen}
\end{align}
The Taylor expansion around the nontrivial fixed point enables us to express $\nu$ in terms of $x_{c}$:
\begin{align}
x_{n} - x_{c} &= x_{n+1}^{u} + x_{n+1}^{v} - x_{c} \notag\\
&\approx x_{c}^{u} + u x_{c}^{u-1}(x_{n+1}-x_{c}) + x_{c}^{v} + v x_{c}^{v-1} (x_{n+1}-x_{c}) - x_{c} \notag \\
&= ( ux_{c}^{u-1} + v x_{c}^{v-1} )(x_{n+1}-x_{c})
\intertext{with}
x_{c}&=x_{c}^{u} + x_{c}^{v}   .\label{eq:fixedPointEq}
\intertext{Feeding this equation into Eq. \eqref{eq:nuTwoGen}, we obtain the final expression as}
\nu &= \frac{\ln(u)}{\ln\left( u x_{c}^{u-1} + v x_{c}^{v-1}\right)}.
\end{align}
Equation (\ref{eq:fixedPointEq}) cannot be solved by hand in general, and hence we must rely on a numerical solver. Exceptional cases will be explained later.

%Simply put, 
%\begin{align}
%\nu = \frac{\ln(u)}{\ln\left( \frac{dx'}{dx}\big|_{x=x_c} \right)}
%\intertext{with RG flow of }
%x'=x^u + x^v
%\end{align}

\section{Existence and uniqueness of a nontrivial fixed point}
In the above argument, we assumed the existence of a positive fixed point $x_{c}$ satisfying Eq. \eqref{eq:fixedPointEq} and the solution $x_{n}$ which meets (\ref{eq:defOfx_n}). 
We prove the existence and the uniqueness of $x_{c}>0$ and that of $x_{n}$ as follows. 

Let us study how the scaling variable $x$ changes under the renormalization-group equation. We define the difference of a scaling variable in the original system and a coarse-grained system as
\begin{align}
f(x):=x^u + x^v -x   .
\end{align}
Because $2\le u \le v$,
\begin{align}
f(0) &= 0,~~~f(1)=1,~~~,f'(0)<0,\\
f''(x) &= u(u-1)x^{u-2} + v(v-1)x^{v-2} > 0 \text{~for~}x>0   .
\end{align}
\begin{figure}
\includegraphics[width=5cm,clip]{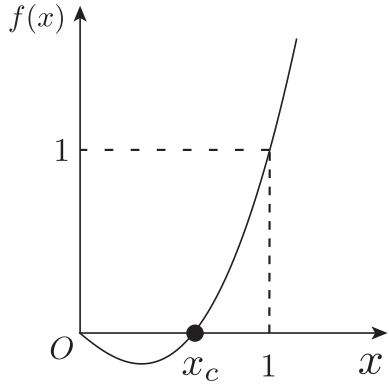}
\centering
\caption{The function $f(x)=x^u+x^v-x$, which has a zero point between $0$ and $1$.}
\label{fig:fx}
\end{figure} 
Therefore, there exists exactly one positive number $x_{c}$ which satisfies $0<x_c<1$ and $f(x_c)=0$ (Figure $\ref{fig:fx}$). In other words, the renormalization-group equation of the self-avoiding walk on the $(u,v)$-flower has exactly one nontrivial fixed point for $2\le u,v$. 
It straightforwardly follows that
\begin{align}
\mu = \frac{1}{x_c} > 1  .
\end{align}
This result is natural, because $\mu$ means the effective coordination number. If $\mu$ were smaller than unity, a walker would quickly come to a dead end and a path could not spread out.\\

Next, we show  using mathematical induction that $G_{n}(R_{n},x)$ is a monotonically increasing function in $x>0$ for $\forall n\in \mathbb{N}$, and that $G_{n}(R_{n},x)$ satisfies $G_{n}(R_{n},0)=0$ and $G_{n}(R_{n},x_{c})=x_{c}$. 
\begin{enumerate}[(i)]
\item $n=1$\\
We have
\begin{align}
\frac{dG_{1}}{dx}(R_{1},x) &= ux^{u-1} + vx^{v-1} > 0 \text{~for~} x>0,\\
G_{1}(R_{1},x_{c}) &= x_{c}^{u} + x_{c}^{v} = x_{c} ,\\
G_{1}(R_{1},0 ) &= 0  .
\end{align}

\item Suppose that the statement is true for $G_{n}(R_{n},x)$. \\
We first prove the monotonicity of $G_{n+1}(R_{n+1},x)$, which is given by
\begin{align}
G_{n+1}(R_{n+1},x) = G_{n}(R_{n}, G_{1}(R_{1},x))  .
\end{align}
Since both $G_{1}$ and $G_{n}$ are monotonically increasing functions, 
the composition of $G_{n}$ and $G_{1}$ is also a monotonically increasing function. Furthermore, 
\begin{align}
G_{n+1}(R_{n+1},x_{c}) &= G_{n}(R_{n}, G_{1}(R_{1},x_{c})) = G_{n}(R_{n},x_{c}) = x_{c},\\
G_{n+1}(R_{n+1},0)     &= G_{n}(R_{n}, G_{1}(R_{1},0    )) = G_{n}(R_{n},0) = 0.
\end{align}
Therefore, the statement is also satisfied for $G_{n+1}$. 
\end{enumerate}
$\whiteBox$

Together with the continuity of $G_{n}(R_{n},x)$, we now proved the unique existence of $x_{n}\in (0,x_{c})$ which satisfies (\ref{eq:defOfx_n}) for an arbitrary constant $e \in (0,x_{c})$. 

\section{Range of $\nu$}
We can study the range of the critical exponent $\nu$ by using inequalities.

We define $x_c$ as the positive solution of (\ref{eq:fixedPointEq}) from now on:
\begin{align}
x_c^{u-1} + x_c^{v-1}=1,~~~2\le u \le v .
\end{align}
First, we can obtain the upper bound of $\nu$ as
\begin{align}
\nu = \frac{\ln(u)}{\ln(ux_{c}^{u-1} + vx_{c}^{v-1}) } 
\le \frac{\ln(u)}{\ln(ux_{c}^{u-1} + ux_{c}^{v-1}) } 
= \frac{\ln(u)}{\ln \left(u \left(x_{c}^{u-1} + x_{c}^{v-1}\right) \right)}=1.
\end{align}
The equality holds iif $u=v$.

We next bound $\nu$ from below. Since $0<x_c<1$ and $u\le v$, we have
\begin{align}
\nu\ge \frac{\ln(u)}{\ln(ux_{c}^{u-1} + vx_{c}^{u-1}) } 
= \frac{\ln(u)} {\ln(u+v) + (u-1)\ln(x_c)} > \frac{\ln(u)}{\ln(u+v)} > 0 .
\end{align}
Furthermore, by setting $x_c = y_{c}^{1/(v-1)}$,
\begin{align}
& y_c < 1,\\
& y_{c}^{\frac{u-1}{v-1}} + y_c = 1,\\
& \lim_{\substack{v\to\infty \\ u:\text{~fixed}}  } y_c = 1,\\
& \lim_{\substack{v\to\infty \\ u:\text{~fixed}}  } y_{c}^{\frac{1}{v-1}} = 1^0 = 1,\\
\intertext{and therefore}
\lim_{\substack{v\to\infty \\ u:\text{~fixed}}  }x_c &= 1.
\end{align}
\begin{equation}
\lim_{\substack{v\to\infty \\ u:\text{~fixed}} } \nu = 
\lim_{\substack{v\to\infty \\ u:\text{~fixed}}  } \frac{\ln(u)}{\ln(ux_{c}^{u-1} + vx_{c}^{v-1}) } = 0   .
\end{equation}
In conclusion, the range of $\nu$ is $0 < \nu \le 1$, and $\nu=1$ holds true iif $u=v$, and $\nu$ can become arbitrarily close to $0$.

\section{Exact results}
As we noted previously, the solution of (\ref{eq:fixedPointEq}) cannot be written down explicitly in general. 
There are, however, exceptional cases where we can obtain $x_c$, $\mu$, and $\nu$ explicitly.

First for the $(u,u)$-flower,
Eq. (\ref{eq:fixedPointEq}) reduces to
\begin{align}
x_c = 2 x_{c}^u \iff x_c = 2^{-\frac{1}{u-1} },
\end{align}
from which we obtain
\begin{align}
\mu &= 2^{\frac{1}{u-1} } ,\label{eq:mu_RG_u_u} \\
\nu &= \frac{\ln(u)}{\ln(u)} = 1  .
\end{align}

Next for the $(u,2u-1)$-flower,
by setting $y = x_{c}^{u-1}$, Eq. (\ref{eq:fixedPointEq}) is reduced to the quadratic equation
\begin{align}
& y^2 + y - 1 = 0, \\
\intertext{which yields}
& y = \frac{-1 + \sqrt{5}}{2} \\
\intertext{because $y>0$, and then}
& x_c = \left( \frac{-1 + \sqrt{5}}{2}  \right)^{\frac{1}{u-1}}  .
\end{align}
We thereby obtain
\begin{align}
& \mu = \frac{1}{x_c} = \left( \frac{-1 + \sqrt{5}}{2}  \right)^{\frac{-1}{u-1}} ,\label{eq:mu_RG_u_2um1} \\
& \nu = \frac{\ln(u)}{ \ln\left( \frac{5-\sqrt{5}}{2}u + \frac{-3 + \sqrt{5}}{2} \right)}  .
\end{align}
In this case, $\nu$ is a monotonically increasing function of $u$, and converges to unity in the limit of $u\to\infty$.

\section{Comparison to the mean-field theory}
Let us compare our analytic expressions with mean-field results. 
A tree approximation is usually referred to as a mean-field theory when we discuss stochastic processes on complex networks.
\footnote{Flory's approximation of $\nu$ is also called a mean-field theory. Readers should not confuse them.}
Under a mean-field approximation, the $(u,v)$-flower is approximated with a tree whose nodes have the same degree as the mean degree of the original flower.

The self-avoiding walk on this tree is identical with the random walk with an immediate return being forbidden (namely, the non-reversal random walk)~\cite{herrero03,herrero05}. Since the connective constant $\mu$ is the effective coordination number,
the tree approximation is 
\begin{align}
\mu = \bra k\ket - 1 = \frac{2M_n}{N_n} - 1 \xrightarrow{n\to\infty}  \frac{u+v}{u+v-2}  .\label{eq:mu_tree}
\end{align}
%A critical exponent $\nu$ is always $\nu=1$ regardless of the values of $u$ and $v$, because the shortest distance between a start point and an end point is identical with the path length.

\begin{figure}
\includegraphics[width=15cm,clip]{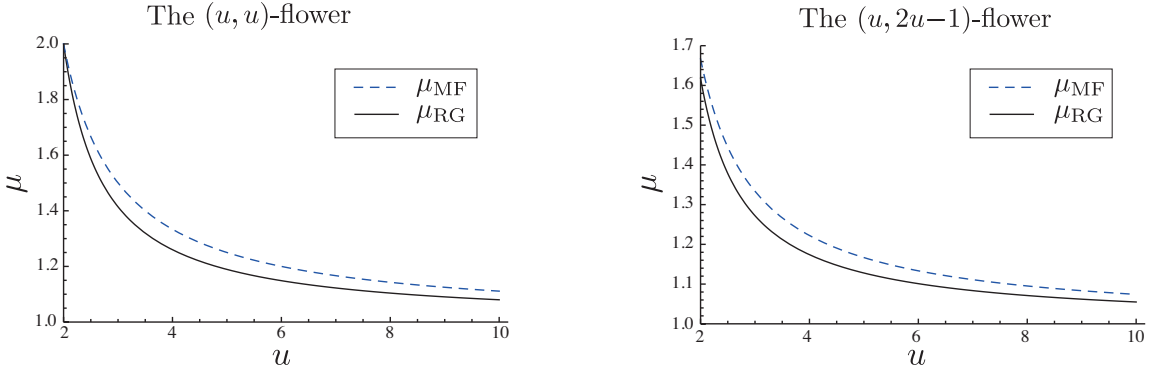}
\centering
\caption{Comparison of the connective constants in the mean-field theory and the renormalization-group. The mean-field estimate $\mu_{\mathrm{MF}}$ is overestimated because loops are ignored.}
\label{fig:compareRGMF_2figs}
\end{figure}

In the mean-field theory, we approximate graphs as trees neglecting loops. The connective constant $\mu$ in the mean-field theory is therefore expected to be overestimated because a walker may encounter a visited site on a graph with loops, and hence the effective coordination number $\mu$ becomes smaller compared to a tree with the same average degree.  We confirm that our expectation is correct both analytically and by numerical calculations. In this section, we first show analytic results (Figure $\ref{fig:compareRGMF_2figs}$).
Numerical simulation will be explained in the next chapter.

First for the $(u,u)$-flower, 
we obtain from (\ref{eq:mu_RG_u_u}) and (\ref{eq:mu_tree})
\begin{align}
\mu_{\mathrm{RG}} &= 2^{\frac{1}{u-1}},\\
\mu_{\mathrm{MF}} &= \lim_{n\to\infty} \bra k \ket - 1=\lim_{n\to\infty} \frac{2M_n}{N_n} - 1 = 1 + \frac{1}{u-1},
\intertext{which means}
\mu_{\mathrm{MF}}& \ge \mu_{\mathrm{RG}}  .
\end{align}

Second for the $(u,2u-1)$-flower,
we obtain from (\ref{eq:mu_RG_u_2um1}) and (\ref{eq:mu_tree}) the following:
\begin{align}
\mu_{\mathrm{RG}} = \left( \frac{\sqrt{5}+1}{2} \right) ^{\frac{1}{u-1}},\\
\mu_{\mathrm{MF}} = 1 + \frac{2}{3u-3}  .
\end{align}
We can show $\mu_{\mathrm{MF}} > \mu_{\mathrm{RG}}$ by setting $z = 1/(u-1)$ with $0<z\le 1$.

\section{Exact solution}
We can obtain the exact solution for the $(u,u)$-flower without relying on the renormalization-group analysis in Sec. \ref{sec:RG}.
Using (\ref{eq:recursionG_n}) repeatedly, we obtain
\begin{align}
G_1(R_1,x) &= x^u + x^u = 2x^u ,\\
G_2(R_2,x) &= G_1(R_1,x) ^u + G_1(R_1,x) ^u = 2 G_1(R_1,x) ^u = 2 (2x^u)^u \nonumber \\
           &= 2^{u+1}x^{u^2} ,\\
G_3(R_3,x) &= G_2(R_2,x) ^u + G_2(R_2,x) ^u = 2 G_2(R_2,x) ^u = 2 ( 2^{u+1}x^{u^2} )^u \nonumber\\
           &= 2^{u^2 + u+1}x^{u^3} ,\\
\cdots\\
G_n(R_n,x) &= 2^{u^{n-1} + u^{n-2}+\cdots + 1} = 2^{\frac{u^n-1}{u-1} }x^{u^n} ,
\end{align}
which are cast into the form
\begin{align}
\exp\left( -\frac{R_n}{\xi(x)} \right) &= G_n(R_n,x) = 2^{\frac{u^n-1}{u-1} }x^{u^n},
\intertext{with}
\xi(x) &= -\frac{R_n}{\ln\left( 2^{\frac{u^n-1}{u-1} }x^{u^n} \right)} = -\frac{u^n}{\ln\left( 2^{\frac{u^n-1}{u-1} }x^{u^n} \right)}.
\end{align}
Let $x_{c}^{(n)}$ be
\begin{align}
x_{c}^{(n)}:=2^{\frac{-1+u^{-n}}{u-1}}  .
\end{align}
We then have $0<\xi(x) < \infty$ when $0<x<x_{c}^{(n)}$ and $\xi(x)$ diverges as $x\nearrow x_{c}^{(n)}$.
The Taylor expansion around $x_{c}^{(n)}$ gives
\begin{align}
\xi(x) = \frac{2^{ \frac{-1+u^{-n}}{u-1} }}{x_{c}^{(n)} - x + O((x_{c}^{(n)} - x)^2)}.
\end{align}
In the thermodynamic limit, we arrive at 
\begin{align}
&\lim_{n\to\infty}x_{c}^{(n)} = 2^{\frac{-1}{u-1}}=:x_c ,\\
&\xi(x)\xrightarrow{n\to\infty} = \frac{2^{\frac{-1}{u-1}}}{x_c - x + O((x_c-x)^2)} .
\end{align}
The critical point $x_{c}^{(n)}$ is shifted from $x_{c}$ because of a finite-size effect. This effect disappears when the system size becomes infinite and the critical point reaches the correct value in the thermodynamic limit. Finite-size effects are often observed in numerical simulation. Quantities which should ideally diverge become smooth functions and the divergence is never observed in computers. In our case, the correlation function {\it actually diverges} even in a finite system.\\

%!!!!WRITE HERE THE CONSISTENSY WITH DIMENSIONAL ANALYSIS OF FINITE SIZE EFFECT TAUGHT BY HATANO-SENSEI!!!!

In this section, we have rigorously proved the following theorem. \\

%\begin{screen}
The critical exponent $\nu$ of the self-avoiding walk on the $(u,u)$-flower is $\nu=1$. 
%\end{screen}\\
The fractal dimension of the $(u,u)$-flower is $d_{\mathrm{f}} = \ln (2u) / \ln(u)$,
 which takes a value $1<d_{\mathrm{f}}\le 2$. Therefore, the following corollary holds:\\
 
%\begin{screen}
There is no one-to-one correspondence between the fractal dimension and the critical exponent $\nu$. 
%\end{screen}\\
Note that it is conjectured that $\nu = 3/4$ in $\mathbb{R}^2$.

As we mentioned in Chapter 1, the critical exponents of the self-avoiding walk in the Euclidian space are determined only by the dimensionality. 
Extension of the self-avoiding walk from the Euclidian space to fractals  increases the number of universality classes from finite to infinite. 

\chapter{Numerical Simulation}
In Chapter 2, we used some hypotheses to derive the connective constant $\mu$ and the critical exponent $\nu$. 
In order to confirm the hypotheses, we here present numerical simulations. 

Furthermore, we delve into the exponent $\nu$ in the context of diffusion. 
We observe that the mean shortest distance between the starting point and the end point increases as $\overline{d_{k}^{(s)}} \approx k^{\nu'}$, where $k$ is the length of a path, $\overline{d_{k}^{(s)}}$ is the mean shortest distance, and $\nu'$ is an exponent which describes the speed of diffusion. 
We numerically show that the exponents $\nu'$ and $\nu$, which was defined through the two-point function, are the same: $\nu'=\nu$.  

Numerical simulations of the self-avoiding walk are divided into two major categories: enumeration algorithms and Monte-Carlo methods. Enumeration algorithms count the number of paths with no approximation while Monte-Carlo methods count the number of paths or measure distances from a starting point by generating random numbers. We used the depth-limited search, which is a modification of the depth-first search and is the most straightforward enumeration algorithm of self-avoiding paths.
We adopted a biased sampling method as a  Monte-Carlo method.

\section{Ensemble of fixed length paths}
Before going to describe algorithms, let us define an ensemble first. 
Let $G=(V,E)$ be a connected finite graph. 
A self-avoiding path of length $k$ is defined as 
\begin{align}
&\omega = (\omega_0, \omega_1,\cdots, \omega_k),~~~\omega_i \in V,\\
&\omega_i\neq\omega_j \text{~for~} i\neq j,\\
&(\omega_i,\omega_{i+1})\in E  .
\end{align}
The set of paths of length $k$ for a fixed start point and a free end point is
\begin{align}
\Omega_{k}^{(s)} := \text{(All self-avoiding paths of length $k$ which start from $s$)} .
\end{align}

In order to discuss the speed of diffusion of a graph later,  we define a distance on a graph here. For any nodes $v_1,v_2\in V$,
\begin{align}
d(v_1,v_2) := \text{(The shortest distance between $v_1$ and $v_2$)}. \label{eq:shortestDist}
\end{align}
We can easily confirm that Eq. \eqref{eq:shortestDist} satisfies the axioms of norm.\footnote{The shortest distance is also called the chemical distance. Especially on $\mathbb{Z}^2$, it is called the Manhattan distance.} 

In order to consider the typical end-to-end distance, we have to define a probability distribution. 
Because we are now considering a finite graph, the number of paths on it is finite, and therefore 
we can introduce a uniform measure without confusion. 
Fixing a path length $k$ and a start node $s$, we introduce a probability measure such that
\begin{align}
P(\omega) = \frac{1}{\#\Omega_{k}^{(s)}}~~~~\forall \omega \in \Omega_{k}^{(s)}  . \label{eq:uniformMeasure}
\end{align}
Though this is not a serious problem, we attention the reader that when the graph is too small, it may not contain a path of length $k$ and $\#\Omega_{k}^{(s)}=0$.

The mean shortest distance of paths of which length is $k$ and which start from node $s$ is given by
\begin{align}
\overline{d_{k}^{(s)}}:=\frac{ \sum_{(\omega_0,\omega_1,\cdots,\omega_k)\in\Omega_{k}^{(s)} }d(\omega_0,\omega_k) }{\#\Omega_{k}^{(s)}}   , \label{eq:MSD}
\end{align}
where we took the average over the uniform distribution \eqref{eq:uniformMeasure}.
We define the exponent of displacement $\nu'$ as
\begin{align}
\overline{d_{k}^{(s)}} \approx k^{\nu'}  .  \label{eq:nuPrimeDef}
\end{align}
In order to distinguish $\nu'$ from $\nu$, which is defined in Eq. \eqref{eq:nuDef} through the two-point function, we used the prime sign. 

\section{Depth-limited search}
\begin{figure}
\includegraphics[width=4cm,clip]{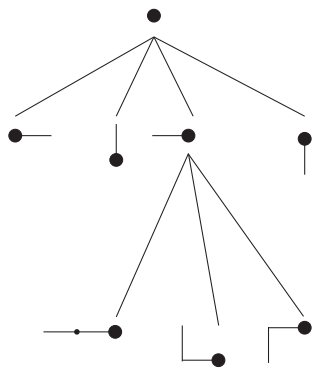}
\centering
\caption{The tree which is explored in the depth-limited search. 
Enumeration of all the paths on the square lattice is considered in the figure. 
The black dot indicates the starting point. 
Nodes of depth $k$  consist of all self-avoiding paths of length $k$ starting from node $s$. Two nodes are connected if the path of the child node can be generated by appending an edge to that of the parent node. }
\label{fig:DLSLattice}
\end{figure}

All the paths of length $k$ can be enumerated using the depth-limited search (DLS). 
In the depth-limited search, we first define a  tree of height $k_{\text{max}}$, whose node represents a self-avoiding path, and next explore the tree by a usual depth-first search (Figure \ref{fig:DLSLattice})~\cite{cormen09}. 
Nodes of depth $k$  consist of all self-avoiding paths of length $k$ starting from node $s$. Two nodes are connected if the path of the child node can be generated by appending an edge to that of the parent node. 
We implemented the depth-limited search using recursion.\footnote{The depth-limited search can be also implemented by using a stack. We implemented the depth-limited search in both ways in C++ and found that the implementation with a stack was slower than the one with recursion, though the latter implementation is accompanied by an overhead of recursion. This is probably because optimization in the one with recursion was carried out efficiently by a compiler.}

%Every node is represented by a class with member variables $color, depth, children$. $color$ is $WHITE$ or $BLACK$, which represents not-found-yet or found respectively. $depth$ is the length of the corresponding SAW path. $children$ is a list of all the children of the node. 

\begin{algorithm}[h]                      
\caption{DLS-Enumeration}         
\label{DLS-Enumeration}                          
\begin{algorithmic}[1]                  
% \STATE Let $Z$ be an array of integers of length $k_{\text{max}}$. $Z[k]$ is the number of SAW paths of length $k$.
\STATE Let $s$ be a node from which a search starts.
\STATE Let $k_{\text{max}}$ be the maximum length of self-avoiding paths.
\STATE $path$ = an array which contains only the start node $s$
\STATE call DLS-Visit($path$, 0)
\end{algorithmic}
\end{algorithm}

\begin{algorithm}[h]                      
\caption{DLS-Visit($path, k$)}         
\label{DLS-Visit}                          
\begin{algorithmic}[1]                  
\IF { $k == k_{\text{max} }$ }
	\STATE return
\ELSE
	\STATE $adjNodes=$ a set of adjacent nodes of the last node.
	\IF { $adjNodes$ is empty }
		\STATE return
	\ELSE
		\FOR {Node $w\in adjNodes$ }
			\STATE Appending $w$ to $path$, create a new path $path'$.
			\STATE $k=k+1$
			\STATE call DLS-Visit($path',k$)
		\ENDFOR 
	\ENDIF
\ENDIF
\end{algorithmic}
\end{algorithm}

The procedure DLS-Enumeration just initializes parameters and calls the procedure DLS-Visit. 
The procedure DLS-Visit works as follows. 
It uses recursion and stops recursion calls if the path length reaches $k_{\text{max}}$ (Line 1). 
If the path length is shorter than $k_{\text{max}}$,  Line 4 defines a set consisting of all the adjacent nodes of the last node. 
If the end point is surrounded by visited sites and has no node to go next,  the path is abandoned (Lines 5 and 6). 
If the end point has at least one adjacent unvisited node, we create the same number of new paths as that of the adjacent unvisited sites and do recursive calls (for-loop 8-12). 

\section{Biased sampling}
\begin{figure}
\includegraphics[width=5cm,clip]{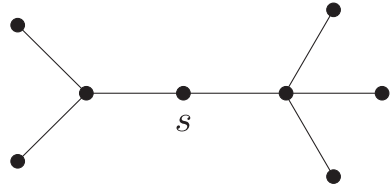}
\centering
\caption{An example of a non-uniform probability distribution of paths. Let us consider a path of length $3$ starting from the node $s$. Making a walker do the random walk and adopting only paths which satisfy the self-avoiding condition, we can obtain a self-avoiding path of length $3$. This random path, however, does not follow a uniform distribution (\ref{eq:uniformMeasure}) because a path which goes to the right appears with the probability $1/6$, while a path to the left does with the probability $1/4$.}
\label{fig:BScounterexample}
\end{figure}

Let us consider how to calculate the mean-shortest distance (\ref{eq:MSD}). 
For this purpose, Monte-Carlo methods are suitable because they can generate longer paths. 
If we can generate every path $\omega\in\Omega_{k}^{(s)}$ with the same probability, 
we will be able to approximate Eq. (\ref{eq:MSD}) by
\begin{align}
\overline{d_{k}^{(s)}} \approx \frac{1}{M} \left(d\left(\omega_{0}^{(1)}, \omega_{k}^{(1)}\right) + d\left(\omega_{0}^{(2)}, \omega_{k}^{(2)}\right) + \cdots + d\left(\omega_{0}^{(M)}, \omega_{k}^{(M)}\right)\right)  .
\end{align}
Here $\omega^{(i)}$ for $1\le i\le M$ is a random variable which follows the distribution (\ref{eq:uniformMeasure}). 
If every node is equivalent, such a path can be created by conducting the random walk. 
For example, on a square lattice, we perform the random walk and accept a path only if it satisfies the self-avoiding condition~\cite{sokal96, binder10}. When all nodes are not equivalent, however, we cannot produce a path which follows a uniform distribution (\ref{eq:uniformMeasure}) by simply letting a walker to choose the next node with an equal probability. An example of a non-uniform measure is  illustrated in Figure \ref{fig:BScounterexample}. In order to calculate the mean shortest distance (\ref{eq:MSD}), we have to take account of weights of paths.

Let $P(\omega)$ be a uniform distribution on a sample space $\Omega_{k}^{(s)}$ and $P'(\omega)$ be any other distribution on $\Omega_{k}^{(s)}$. Denoting the accumulate distribution of $P(\omega)$ and $P(\omega')$ by $\mu$ and $\mu'$ respectively, we can express the average of any quantity $A$ in the form ~\cite{binder10}
\begin{align}
\bra A \ket  
=\frac{\int_{\Omega_{k}^{(s)}} A(\omega)d\mu}{\int_{\Omega_{k}^{(s)}} d\mu} 
= \frac{\int_{\Omega_{k}^{(s)}} A(\omega)P(\omega)d\omega}{\int_{\Omega_{k}^{(s)}} P(\omega)d\omega} 
= \frac{\int_{\Omega_{k}^{(s)}} A(\omega)P(\omega)/P'(\omega)d\mu'}{\int_{\Omega_{k}^{(s)}} P(\omega)/P'(\omega)d\mu'}  .  \label{eq:BSformula}
\end{align}
Therefore, the average over $P(\omega)$ can be calculated by taking the average over an arbitrary distribution $P'(\omega)$ with a weight $P(\omega)/P'(\omega)$. 

The simplest choice of $P'(\omega)$ is the following one. 
Let a walker select the next site randomly among adjacent unvisited sites, 
and we thereby define $P'(\omega)$ as a distribution that the trajectory of the walker follows. 

Let $l_i$ be the number of sites to which a walker can go next in the step $i$. Then a path appears with a probability proportional to $1/\prod_i l_i$. Such a path comes out with a probability
\begin{align}
P'(\omega) = \frac{W}{\prod_{i=0}^{k-1}l_i(\omega) }  ,
\end{align}
where $W$ is a normalization factor. Since $P(\omega)$ and $W$ are constants, these terms in the denominator and numerator of (\ref{eq:BSformula}) cancel each other, and we obtain
\begin{align}
\bra A\ket &= \frac{\int_{\Omega_{k}^{(s)}} A(\omega)\prod_{i=0}^{k-1}l_i(\omega) d\mu'}{\int_{\Omega_{k}^{(s)}} \prod_{i=0}^{k-1}l_i(\omega) d\mu'} \\
&\approx \frac{ A(\omega^{(1)})\prod_{i=0}^{k-1}l_i(\omega^{(1)})  +   \cdots + A(\omega^{(M)})\prod_{i=0}^{k-1}l_i(\omega^{(M)})}{\prod_{i=0}^{k-1}l_i(\omega^{(1)})  + \cdots + \prod_{i=0}^{k-1}l_i(\omega^{(M)})}  .  \label{eq:averageA_BS}
\end{align}
Here $\omega^{(i)}$ for $1\le i\le M$ is a random variable (path) which follows the distribution $P'(\omega)$.

\begin{algorithm}[h]                      
\caption{Biased-Sampling(k)}         
\label{Biased-Sampling}                          
\begin{algorithmic}[1]                  
\STATE $i\_config=0$
\WHILE {$i\_config  <  num\_config$}
	\STATE $path$ = an array which only contains the start node $s$
	\STATE $v=s$
	\STATE $k = 0$
	\STATE $weight = 1$
	\WHILE {$k < k_{\text{max}}$}
		\STATE $adjNodes=$ a set of unvisited nodes adjacent to node $v$
		\IF {$adjNodes$ is empty}
			\STATE break
		\ELSE
			\STATE $w=$a randomly chosen node from $adjNodes$
			\STATE Append $w$ to $path$.
			\STATE $weight = weight \times adjNodes.size$
			\STATE $k=k+1$
			\STATE $v=w$
		\ENDIF
	\ENDWHILE
	\IF {$path.length==k_{\text{max}}$}
		\STATE $i\_config = i\_config + 1$
	\ENDIF
\ENDWHILE
\STATE Accumulate results using (\ref{eq:averageA_BS})
\end{algorithmic}
\end{algorithm}
Let us consider the case where we want to sample $num\_config$ pieces of configurations of length $k\_max$. 
In the pseudocode of Biased-Sampling, $i\_config$ counts the number of realized paths of length $k_{\text{max}}$, and $k$ is the length of a path. $adjNodes.size$ is the number of options to go to next, and hence a weight $weight$ is multiplied by $adjNodes.size$ in the for-loop. The variable $v$ denotes the end point, or a node which is appended in the previous step. 
Paths are created until $num\_config$ pieces of paths are generated (Line 2). 
Lines 4-6 initialize the path and the weight. 
While the path length is shorter than the intended length, a randomly chosen adjacent node of the end point is appended to the current path (Lines 12 and 13). 
However if the end point is surrounded by visited nodes and the path cannot be extended, the path is abandoned and a new path is created (Lines 9 and 10). 
Line 14 multiplies $weight$ by the number of unvisited adjacent nodes. 
Lines 15-16 update the path length $k$ and the end point $v$, respectively. 
Line 20 increments the number of realized configuration if the path length achieves the desired length. 
After $num\_config$ paths of length $k_{\text{max}}$ are generated, 
Line 23 calculates various statistics using Eq. (\ref{eq:averageA_BS}).

\section{The number of paths}
\begin{figure}
\includegraphics[width=13cm,clip]{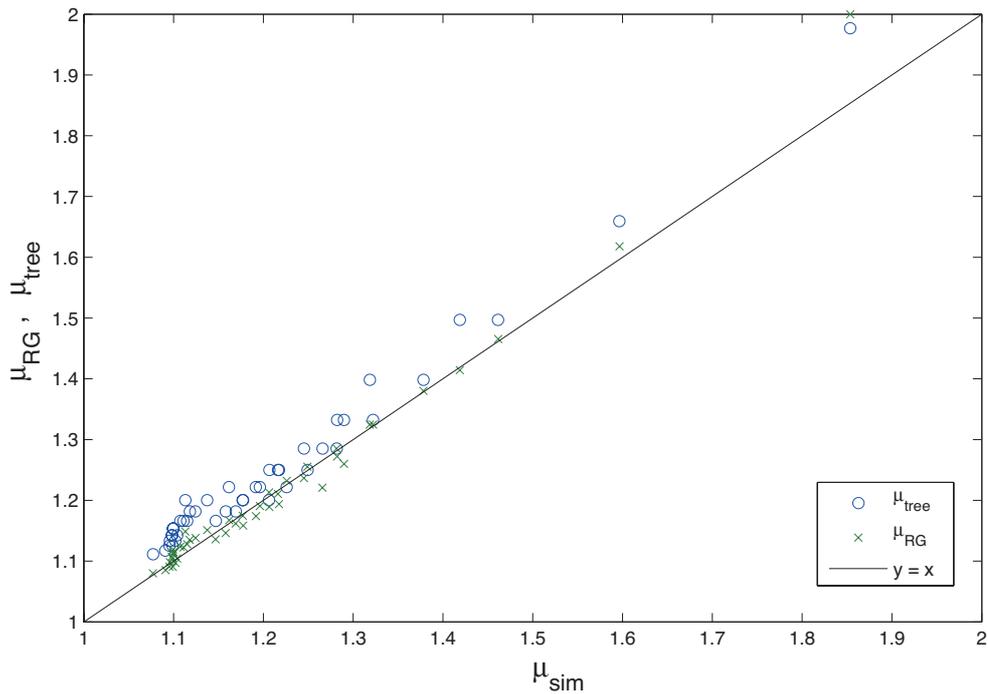}
\centering
\caption{Comparison of the connective constant $\mu$ obtained by three different methods for various $u$ and $v$. The horizontal axis is the estimate of $\mu$ in simulation with the fitting in (\ref{eq:fitting_mu}), while the vertical axis is that of the renormalization-group analysis or the tree approximation (mean-field theory). The simulation condition is $n=4$, $2\le u\le v\le 10,\text{~and~}1\le k\le 30$. A hub was chosen as the starting point.}
\label{fig:compare_mu}
\end{figure}

We computed the number of paths of length $k$ using the depth-first search. Drawing an analogy to Eq. \eqref{eq:CkEuclidian} in the Euclidian spaces, we assume that the number of paths of length $k$ starting from a node $s$ behaves as 
\begin{align}
C_{k}^{(s)} = A^{(s)} \mu^k k^{\gamma-1}  .  \label{eq:C_k}
\end{align}  
Because $C_{k}^{(s)}$ increases exponentially, it is easy to acquire the value of $\mu$, but unfortunately $\gamma$, which is of more interest from the perspective of critical phenomena, is difficult to obtain accurately. 
Choosing a node with the largest degree, namely a hub, as a starting point $s$, we computed $C_{k}^{(s)}$ for $n=4,2\le u\le v\le 10,1\le k\le 30$ and fitted the series $C_{k}^{(s)}$ to
\begin{align}
\ln C_{k}^{(s)} = A' + k\ln\mu + (\gamma -1)\ln k  .  \label{eq:fitting_mu}
\end{align}
We obtained only $\mu$ in high precision (Figure \ref{fig:compare_mu}). As we have noted, the value $\mu$ under the tree approximation is overestimated because the existence of loops is not taken into consideration.  The computational complexity of DLS-Enumeration is roughly $\propto\mu^k$, and hence more running time is needed as $\mu$ gets larger. 

\begin{figure}
\includegraphics[width=4cm,clip]{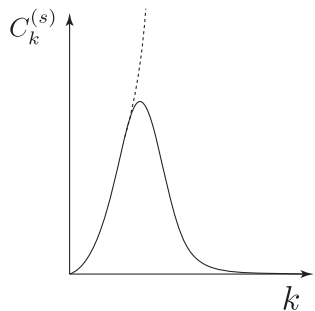}
\centering
\caption{A sketch of the number of paths against the path length. The number of paths increases as (\ref{eq:C_k}) when $k$ is moderately large, and then it starts decreasing due to the finite-size effect. What we need is the asymptotic behavior in the thermodynamic limit indicated by the dashed curve.}
\label{fig:nthFlower_C_k}
\end{figure}

The upper right points of Figure \ref{fig:compare_mu} which correspond to the $(2,2)$-flower deviates from the line. 
This is because the graph is smallest for $(u,v)=(2,2)$ and the finite size effect appears strongly. 
The number of paths first increases and then starts decreasing due to a finite size effect (Figure \ref{fig:nthFlower_C_k}). 
What we need to obtain is the asymptotic behavior of the rise in the intermediate region.

\section{The exponent of displacement $\nu'$}
We hypothesized that the mean shortest distance from the starting point increases as a power function of the path length ($\ref{eq:nuPrimeDef}$). 
We can indeed confirm it by the enumeration algorithm. 
We found that $\ln \overline{ d_{k}^{(s)} }$ ripples around an asymptotic line and the amplitude gets smaller as $k$ becomes larger.

\begin{figure}
\includegraphics[width=15cm,clip]{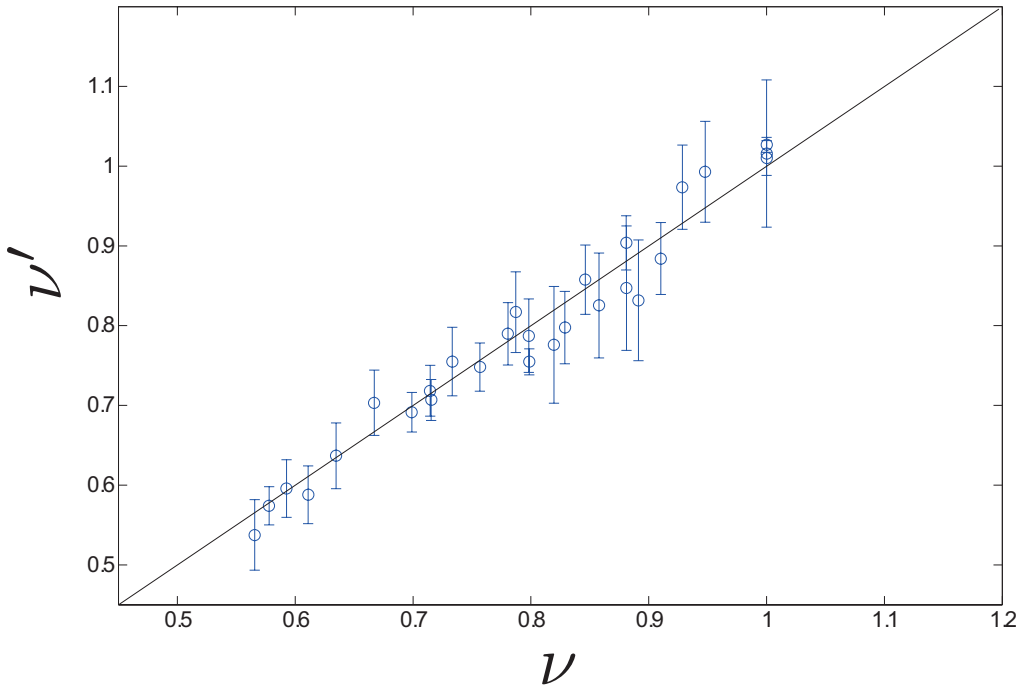}
\centering
\caption{The critical exponent $\nu$ and the exponent of displacement $\nu'$ defined in terms of the shortest distance for various $(u,v)$. This figure supports our hypothesis $\nu=\nu'$. 
We estimate the critical exponent $\nu$ by the renormalization-group analysis and computed the exponent of displacement $\nu'$ by the biased sampling algorithm followed by a curve fitting. We chose a hub as the starting point $s$. 
The simulation condition was $n=4,~2\le u\le 5,$ and $2\le v\le 10$.  
}
\label{fig:nu_nu_MC_bin}
\end{figure}

\begin{figure}
\includegraphics[width=7cm,clip]{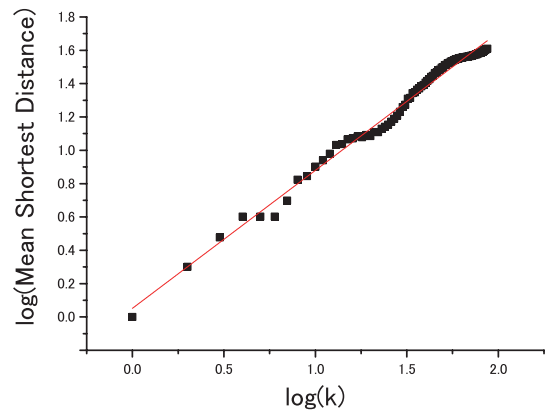}
\centering
\caption{The mean shortest distance $\overline{d_{k}^{(s)}}$ against the path length $k$. The series of $\overline{d_{k}^{(s)}}$ computed by the enumeration algorithm is fitted to $\ln\overline{d_{k}^{(s)}} = A + \nu' \ln k$. We chose a hub was chosen as the starting point $s$. We counted all the paths of $k\le 87$ for $(u,v,n)=(3,5,5)$. The estimate of $\nu'$ is 0.82779, while $\nu$ evaluated by the renormalization-group analysis is 0.828851.}
\label{fig:MeanShortestDistance_k}
\end{figure}

Next, we hypothesized that the exponent $\nu$, which is defined through the generating function by (\ref{eq:nuDef}), is equal to the exponent of displacement $\nu'$ in (\ref{eq:nuPrimeDef}): 
\begin{align}
\nu=\nu'  .
\end{align}
We confirmed this hypothesis by the biased sampling algorithm (Figure $\ref{fig:nu_nu_MC_bin}$). 
The simulation condition is $2\le u\le 5,~2\le v\le 10$, and $num\_config = 10,000$ configurations of paths were generated for each $k$. 
We used a hub as the starting node $s$. 
Assuming the relation $\eqref{eq:nuPrimeDef}$, we fitted the estimate of the obtained mean shortest distance $\overline{d_{k}^{(s)}}$ to
\begin{align}
\ln  \overline{d_{k}^{(s)}}  = A + \nu' \ln k  .   \label{eq:nuPrimeModel}
\end{align}
The maximum path length $k_{\text{max}}$ is the value just before the finite-size effect appears and $\overline{d_{k}^{(s)}}$ starts decreasing. 
An example of fit to \eqref{eq:nuPrimeModel} is shown in Figure \ref{fig:MeanShortestDistance_k}.
We rejected the data point for $(u,v)=(2,2)$ because $k_{\text{max}}$ was too small, 
and fitting could not be done. 
Detail of analysis is written in Appendix \ref{chap:MCAnnal}.

\chapter{Discussion}
Our study has revealed two things. 
The first finding is that consideration of the self-avoiding walk on fractal complex networks leads to better understanding of diffusion in fractals. 
The second discovery is that the scaling theory of polymers is applicable to graphs on which the Euclidian distance is not defined.

The connection between the self-avoiding walk to a spin system in the Euclidian space was first noticed by de Gennes~\cite{deGennes72, deGennes79}. 
During the 1980s and 1990s, many fractal lattices embedded in Euclidian spaces were studied~\cite{dhar78, gefen80, havlin84, lam84, rammal84, friedberg86, havlin87, aharony89, benavraham00, hattori04}. One of the most famous fractal lattices is the Sierpinski gasket. 
The Sierpinski gasket can be generalized to higher dimensions, but analysis of dynamics on them is not easy. 
In fact, the self-avoiding walk on high-dimensional Sierpinski gaskets have been solved only in $d=2,3$. Because graphs do not have the constraint that they must lie on the Euclidian spaces, analysis of dynamics of such fractal networks may become easier. 
Indeed, we derived exact results for the self-avoiding walk on the $(u,u)$-flowers for all $u,v\ge 2$. 

In the Euclid space, it is widely believed that the critical exponents of the self-avoiding walk is a function of only the Euclidian dimension. 
In contrast, generalizing the problem to fractals, it has been conjectured that there is no one-to-one correspondence between the similarity dimension and the universality class of the self-avoiding walk since the 1980s. 
We succeeded to confirm this conjecture rigorously. 
We think that fractal complex networks are of use to understand scaling of other stochastic models too.

Enumeration of simple paths is a classical problem in computer science. 
We believe that it is important to understand the scaling properties of paths on graphs. 
Thus, our exact solutions will be of use to study how fast the number of paths increases as the path length gets larger. 
In addition, we showed that the renormalization-group analysis can predict the speed of increase of the mean shortest distance from a starting point.

Scaling theories have traditionally been exploited in the Euclidian spaces~\cite{cardy96, nishimori11}, but 
recent studies of complex networks have unveiled that it is also useful to understand dynamics on graphs~\cite{song05, yook05, goh06,song06, radicchi08, serrano08, radicchi09, rozenfeld10}. 
In particular, there has been great progress in Markov processes on complex networks. 
The most notable result is the scaling theory of the mean first-passage time of random walks on complex networks~\cite{condamin07, gallos07b, condamin08}. 
They showed that the scaling of the mean first-passage time is determined only by fractal dimensions. 
Their result is also applicable to graphs, on which the distance is measured by the shortest distance. 

In contrast to scaling theories for Markovian processes, those for non-Markovian processes such as the self-avoiding walk are poorly understood. 
The methodology of the renormalization group is also applicable to non-Markovian dynamics on graphs as well as Markovian processes. 
We believe that this direction of research will deepen our understanding of non-Markovian dynamics on complex networks.

It will be interesting to consider the extension of Flory's approximation of $\nu$ to fractal graphs. 
It is impossible to apply the renormalization-group analysis which we developed in this thesis to real complex networks 
because graphs for which the exact renormalization can be applied are limited; 
this is the fundamental limit of the present renormalization-group analysis. 
We, however, expect that the scaling properties of paths on networks are determined only by a few parameters such as fractal dimensions. 
Our exact results will serve as a basic model to develop a scaling theory which is generally applicable. 
For example, we found that the speed of the increase of the mean shortest distance between end-to-end points is related with the critical exponent of the zero-component ferromagnet on the $(u,v)$-flower. 
If $\nu$ can be expressed as a function of a few parameters,  
we can predict the number of the $s-t$ paths and the mean shortest end-to-end distance 
beforehand. 
It is also of interest to study whether the critical exponent $\nu$ of the $n$-vector model in the limit $n\to 0$ is the same as that of the exponent of displacement $\nu'$, which 
is defined by $\overline{d_{k}^{(s)}} \approx k^{\nu'}$ on other fractal networks.

\appendix
\chapter{Additional Results of Numerical Simulation} 
We described all important results in the main chapters, but we here present
several additional results.

\section{The number of paths}
A shortcoming of an enumeration algorithm is that long paths are difficult to sample, though what we need is a quantity for large $k$ in Eq. (\ref{eq:nuPrimeDef}). 

\begin{figure}
\includegraphics[width=16cm,clip]{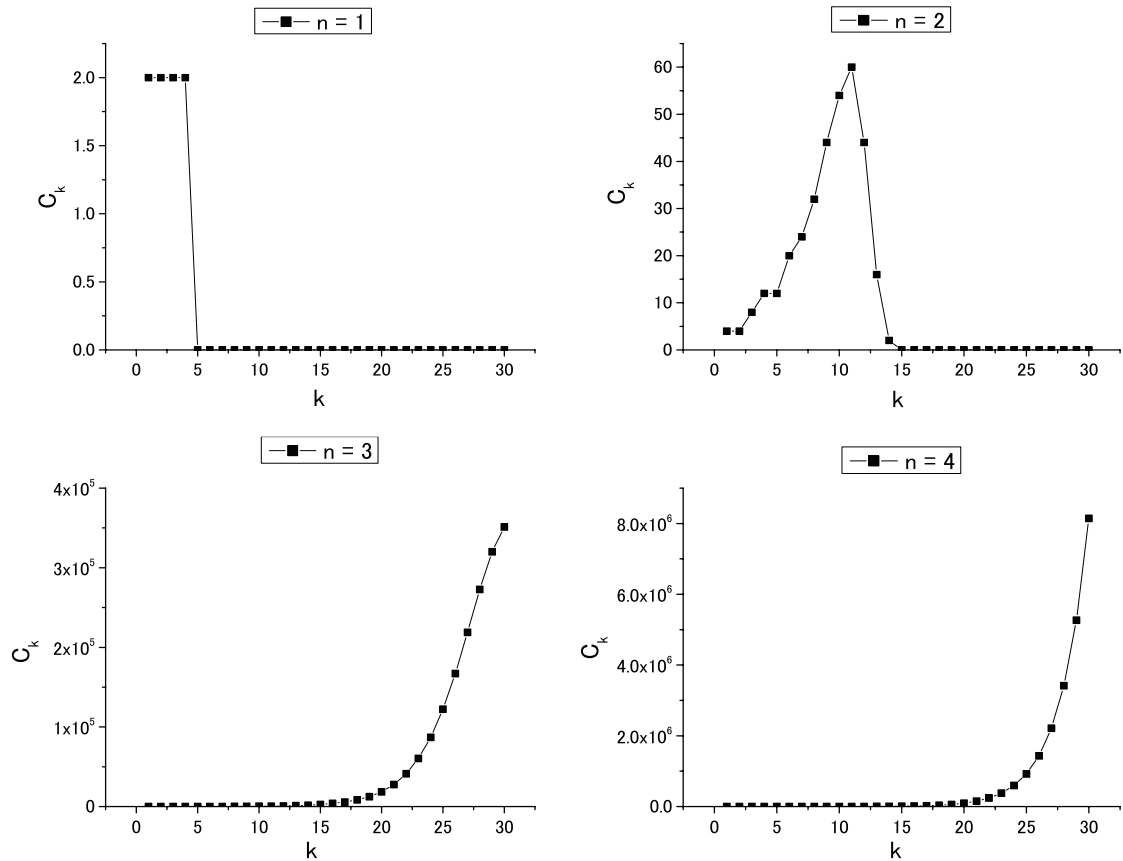}
\centering
\caption{The number of paths of length $k$ for various generations of the $(2,3)$-flower. The number of paths increases as (\ref{eq:C_k}) when $k$ is small, and then it starts decreasing due to a finite-size effect.
}
\label{fig:nth_2_3_Flower}
\end{figure}

As we noted, the number of paths increases as (\ref{eq:C_k}) when $k$ is small, and then it starts decreasing due to a finite-size effect (Figure \ref{fig:nthFlower_C_k}). 
We verified this for various generations of the $(2,3)$-flower (Figure $\ref{fig:nth_2_3_Flower}$).

\section{The exponent of displacement $\nu'$ by the depth-limited search}
\begin{figure}
\includegraphics[width=10cm,clip]{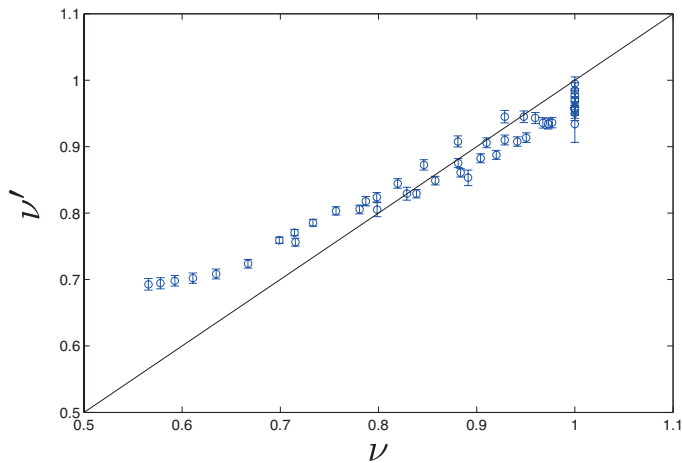}
\centering
\caption{The critical exponent $\nu$ calculated by the renormalization-group analysis and the exponent of displacement $\nu'$ estimated by the enumeration algorithm followed by a curve fitting. 
The points on the left deviate from the line $y=x$ because the mean end-to-end shortest distance $\overline{d_{k}^{(s)}}$ does not still reach an asymptotic region.}
\label{fig:nu_RG_nuPrime_enumeration}
\end{figure}

Longer paths should be obtained in order to determine the exponent $\nu'$ accurately. 
To estimate $\nu'$, therefore, the biased sampling method was more suitable than the depth-limited search. 
We also estimated $\nu'$ by the enumeration algorithm for cross-check (Figure $\ref{fig:nu_RG_nuPrime_enumeration}$). 
The systematic error discussed later is not taken into consideration.

\chapter{Analysis of the Monte-Carlo Simulation} \label{chap:MCAnnal}
In this appendix, we explain the method used in data analysis of estimation of the exponent of displacement $\nu'$ in Figure $\ref{fig:nu_nu_MC_bin}$. 

The mean shortest distance $\overline{d_{k}^{(s)}}$ computed by the biased sampling method is accompanied by an error: 
\begin{align}
\text{(The mean shortest distance from the starting point}) = \overline{d_{k}^{(s)}} \pm \sigma_{k} .
\end{align}
Here $\overline{d_{k}^{(s)}}$ is the sample mean of the shortest distances $d_{k}^{(s)}$ over $num\_config$ realizations 
and $\sigma_{k}$ is the sample standard deviation of $d_{k}^{(s)}$. 

The estimates of $\nu'$ are accompanied by two kinds of errors: a statistical error and a systematic error. 
The statistical error, which is indicated by  $\sigma_{k}$,  comes from fluctuation of random numbers.
The systematic errors, on the other hand, is due to insufficient path lengths in our case; for example, 
\begin{itemize}
\item $\overline{d_{k}^{(s)}}$  has not reached the asymptotic region.
\item $\ln\overline{d_{k}^{(s)}}$ ripples around the asymptotic line. 
\end{itemize}
What we need to obtain is the asymptotic behavior of $\overline{d_{k}^{(s)}}$ as $k\to\infty$. Therefore, $\nu'$ estimated by the fitting $\eqref{eq:nuPrimeModel}$ is not accurate if $k$ is too small. Furthermore, we found that $\ln \overline{ d_{k}^{(s)} }$ ripples around the asymptotic line (Figure \ref{fig:MeanShortestDistance_k}), 
and the amplitude of oscillation gets smaller as $k$ becomes larger. 
In other words,  the model $\eqref{eq:nuPrimeModel}$ is not correct in a strict sense, 
because the average $\overline{d_{k}^{(s)}}$ does not converge to $e^{A}k$ in the limit $num\_config\to\infty$. 
We define $\sigma_{k}'$ as the standard deviation of $\ln \overline{d_{k}^{(s)}}$:
\begin{align}
\sigma_{k}' := [\ln( \overline{d_{k}^{(s)}} + \sigma_{k}) - \ln (\overline{d_{k}^{(s)}} - \sigma_{k}) ]/2. \label{eq:sigmakprime}
\end{align}

Thus, simply fitting $\ln k$ and $\ln\overline{d_{k}^{(s)}}$ with a weight $1 / \sigma_{k}'^{2}$ has two downsides. 
First, because the least-square method minimizes the total of residuals, 
the fitting is done mainly using the data points of small $k$, whose oscillation is large, while the data with large values of $k$ are little taken into consideration. 
Second, the error of the estimate of $\nu'$ is underestimated 
because the systematic error of oscillation is neglected. 
The problem is that we do not know the amplitude of oscillation, and hence 
we resort to  an {\it ad hoc} prescription to take the systematic error into consideration.

We obtain the estimates of $\nu'$ for each $(u,v)$-flower in the following procedure:

\begin{enumerate}
\item Generate $num\_config$ pieces of paths and compute the sample mean shortest distance $\overline{d_{k}^{(s)}}$ and the sample mean standard deviation of $\overline{d_{k}^{(s)}}$, {\it i.e.},  $\sigma_{k}$. 

\item Remove the first data of $\overline{d_{k}^{(s)}}$ whose sample mean standard deviations are zero. 

\item If there still remains $\overline{d_{k}^{(s)}}$ whose sample mean standard deviation is zero, then set $\sigma_{k}$ to the average of the sample mean standard deviations of neighboring data points.
Compute the $\sigma_{k}'$ defined by Eq. \eqref{eq:sigmakprime}

\item Divide the data points into $n_{b}$ bins of the same width in $\ln k$. 

\item Do fitting inside each bin using $\eqref{eq:nuPrimeModel}$ with a weight $1/\sigma_{k}'^{2}$ 
and calculate the root mean square of the residuals of $\ln \overline{ d_{k}^{(s)} }$. 
We denote this root mean square by $s_{i}$. 
Let the mean of $\ln k$ in each bin be $\ln k_{i}$ and that of $\ln \overline{ d_{k}^{(s)} }$ be $\ln \overline{ d_{k_{i}}^{(s)} }$.

\item Fit the data $\ln k_{i}$ and $\ln \overline{ d_{k_{i}}^{(s)} }$ $~~(i=1,\cdots, n_{b})$ in all bins with a weight $1/s_{i}^{2}$ using the model $\eqref{eq:nuPrimeModel}$. 
We denote the error of the estimate of $\nu'$ as $\delta\nu'$.
\end{enumerate} 

Step 2 is intended to remove the region where the distance increases linearly. 
Otherwise $s_{1}$ would become zero in Step 5. 
Step 3 is necessary to fit $\ln \overline{ d_{k}^{(s)} }$ with finite weights in Step 5. 
Steps 5 and 6 are done to coarse-grain data so as to assign a larger weight to the data with larger $k$ and to avoid the underestimation of the error of the estimate of $\nu'$ by taking account of the effect of oscillation as a systematic error. 

Thus obtained estimates of $\nu'$  are plotted in Figure $\ref{fig:nu_nu_MC_bin}$ when the number of bins is $n_{b}=5$. 

%The value of $n_{b}$ is determined so that $\chi^{2}$ is minimized stipulating that each $\nu'$ follows a Gauss distribution with the mean of $\nu$ and the STD of $\delta\nu'_{uv}$. 
%That is, the $\chi^{2}$ is
%\begin{align}
%\chi^{2} = \frac{1}{N} \sum_{u,v} \frac{(\nu' - \nu)^{2}}{ \delta\nu'^{2}_{uv}} 
%\end{align}
%, where $N$ is the number of data points (the degree of freedom). 
%In Figure $\ref{fig:nu_RG_nuPrime_enumeration}$, $\chi^{2} = 22.2,~N=29$.

\bibliographystyle{unsrt}
\bibliography{MastersThesisBib}
\end{document}